\documentclass[a4paper,12pt]{article}

\PassOptionsToPackage{table}{xcolor}
\usepackage{jheppub} 

\usepackage[T1]{fontenc} 

\usepackage[all]{xy}
\usepackage{bbold}
\usepackage[percent]{overpic}
\usepackage{slashed}
\usepackage{wrapfig}
\usepackage{tabu}
\usepackage{diagbox}
\usepackage{mathrsfs,amsmath,amssymb,amsthm,amsfonts,tikz,graphicx,accents,hyperref, color}
\usepackage{dsfont,epiolmec, latexsym, stmaryrd, comment}
\usepackage{slashed,ccaption}
\usepackage{mathrsfs, calligra}
\usepackage{leftidx}
\usepackage{import}
\usepackage{multirow}
\usepackage{amsfonts}
\usepackage{pifont}
\usepackage{tabularx}
\usepackage[utf8]{inputenc}
\usetikzlibrary{intersections,calc}
\usepackage{tikz-3dplot}
\usepackage{ifthen}
\usetikzlibrary{arrows}
\usepackage{amsmath}
\usepackage{xcolor}

\usepackage{caption} 
\DeclareMathOperator{\extdm}{d}
\newcommand{\extd}{\extdm \!}

\usepackage{array}
\usepackage[percent]{overpic}
\usepackage{wrapfig}
\usepackage{bbm}
\usepackage{tabu}
\usepackage{slashed}
\usepackage{fancyhdr} 
\usepackage{amsmath}
\usepackage{amsfonts}
\usepackage{amssymb}
\usepackage{diagbox}

\hypersetup{ linktoc=all,
    colorlinks, linkcolor={blue}, 
    citecolor={red}, urlcolor={darkpink}
}

\graphicspath{{Images/}}

\definecolor{darkred}{rgb}{0.7,0,0}

\definecolor{light gray}{RGB}{220,220,220}
\definecolor{dark purple}{RGB}{108,0,217}
\definecolor{pink}{RGB}{190,20,100}
\definecolor{orang}{RGB}{193,63,0}
\definecolor{green}{RGB}{11,98,17}
\definecolor{darkpink}{RGB}{153,0,76}
\definecolor{bluegreen}{RGB}{0,102,102}
\definecolor{greenlagan}{RGB}{0,102,0}
\definecolor{redgreen}{RGB}{102,102,0}
\definecolor{Redgreen}{RGB}{153,76,0}
\definecolor{vividviolet}{rgb}{0.62, 0.0, 1.0}
\definecolor{amaranth}{rgb}{0.9, 0.17, 0.31}
\definecolor{palatinateblue}{rgb}{0.15, 0.23, 0.89}
\definecolor{brightpink}{rgb}{1.0, 0.0, 0.5}
\definecolor{cornflowerblue}{rgb}{0.39, 0.58, 0.93}
\definecolor{deepcarminepink}{rgb}{0.94, 0.19, 0.22}
\definecolor{radicalred}{rgb}{1.0, 0.21, 0.37}

\usepackage{graphicx}
\usepackage{tikz}
\usetikzlibrary{arrows,chains,shapes,matrix,positioning,scopes}
\usetikzlibrary{decorations.markings}
\tikzstyle arrowstyle=[scale=1]
\tikzstyle directed=[postaction={decorate,decoration={markings,
    mark=at position .65 with {\arrow[arrowstyle]{stealth}}}}]
\tikzstyle reverse directed=[postaction={decorate,decoration={markings,
    mark=at position .65 with {\arrowreversed[arrowstyle]{stealth};}}}]
\usepackage{import}
\usepackage{accents}
\usepackage{mathrsfs,amsmath,amssymb,slashed}
\usepackage{multirow,multicol}
\usepackage{enumitem}

\usepackage[percent]{overpic}
\usepackage{slashed}

\usepackage{wrapfig}
\usepackage{tabu}
\usepackage{tikz}
\usepackage{diagbox}
\usepackage{comment}
\usepackage{tikz}
\usepackage{mathrsfs,amsmath,amssymb,amsthm,amsfonts,graphicx,accents,hyperref,color}
\usepackage{leftidx}
\usepackage{import}
\usetikzlibrary{decorations.pathmorphing}
\DeclareFontFamily{OT1}{rsfs}{}

\DeclareFontShape{OT1}{rsfs}{m}{n}{ <-7> rsfs5 <7-10> rsfs7 <10->rsfs10}{} 

\DeclareMathAlphabet{\mycal}{OT1}{rsfs}{m}{n}

\newcommand{\B}{\mathcal{B}}
\newcommand{\p}{\prime}

\newcommand{\cT}{\mathcal{T}}

\newcommand{\cTi}{\mathcal{T}_{(1)}}

\newcommand{\cPi}{\mathcal{P}_{(1)}}
\newcommand{\cP}{\mathcal{P}}

\newcommand{\laa}{\lambda}

\newcommand{\A}{\mathcal{A}}

\newcommand{\nn}{\nonumber}

\newcommand{\be}{\begin{equation}}
\newcommand{\ee}{\end{equation}}
\usepackage{cancel}

\newcommand{\cA}{\mathcal{A}}
\newcommand{\cF}{\mathcal{F}}

\author{Hamid Afshar$^{\,a,b}$, Erfan Esmaeili$^{\,b}$, H. R. Safari$^{\,b}$}

\title{{\LARGE{Flat space holography in spin-2 extended dilaton-gravity}}}

 \affiliation[a]{\it Department of Physics, Faculty of Science, Ferdowsi University of Mashhad, Mashhad, Iran}
\affiliation[b]{\it  School of Physics, Institute for Research in Fundamental
Sciences (IPM),\\ P.O.Box 19395-5531, Tehran, Iran}

\emailAdd{afshar,erfanili,hrsafari@ipm.ir}

\preprint{IPM/P-2021/002}

\abstract{We present an interacting spin-2 gauge theory coupled to the two-dimensional dilaton-gravity in flat spacetime.
The asymptotic symmetry group is enhanced to the central extension of Diff$(S^1)\ltimes C^\infty(S^1)\ltimes$Vec($S^1$) when the central element of the
Heisenberg subgroup is zero (vanishing $U(1)$ level). Using the BF-formulation of the model we derive the corresponding boundary coadjoint action which is the spin-2 extension of the warped Schwarzian theory at vanishing $U(1)$ level. We also discuss the thermodynamics of black holes in this model.
}

\begin{document}
\maketitle

\section{Introduction}

Two-dimensional spacetimes with asymptotic boundaries have been recently a desirable playground for holographic studies (see e.g. \cite{Almheiri:2014cka,Maldacena:2016upp,Jensen:2016pah,Engelsoy:2016xyb,Saad:2018bqo,Saad:2019lba}). The holographic relationship connects a certain 2D dilaton-gravity model in the bulk to a 1D quantum statistical model in the boundary. A famous example is the  
 Sachdev--Ye--Kitaev (SYK) model \cite{Kitaev:15ur,Sachdev:1992fk,Sachdev:2010um,Kitaev:2017awl}, a solvable  quantum statistical mechanics which in its low energy  has a holographic description in terms of the Jackiw–Teitelboim (JT) \cite{DHoker:1982wmk,Teitelboim:1983ux,Jackiw:1984je,Jackiw:1984,Teitelboim:1984} gravity with nearly AdS$_2$ boundary conditions \cite{Maldacena:2016hyu}  describing the near-horizon geometry of nearly extremal black holes \cite{Mandal:1991tz,Witten:1991yr,Nayak:2018qej}.  The SYK model possesses 1D conformal invariance i.e. the full reparametrization \text{Diff}($S^1$) in its infrared, however, this is spontaneously broken to SL(2,$\mathbb{R}$), and the broken phase is parametrized by Goldstone modes living on the quotient space \text{Diff}($S^1$)/SL(2,$\mathbb{R}$). The effective action for the associated Goldstone modes has an effective description in terms of the Schwarzian action \cite{Kitaev:15ur,Maldacena:2016hyu} which itself is obtained as the boundary action in JT gravity upon imposing appropriate boundary conditions (see e.g.~\cite{Grumiller:2015vaa,Maldacena:2016upp,Cvetic:2016eiv,Davison:2016ngz,Grumiller:2017qao,Mertens:2018fds}).

It is certainly motivating to go beyond JT gravity and try to put other 2D dilaton-gravity models under a holographic test. An important link in this holographic setup is the role played by the Schwarzian action. On the quantum mechanics side, it arises in the large $N$ and strong coupling limit of a SYK-like model and on the gravity side after imposing suitable boundary conditions and as a specific boundary term in a JT-like gravity model. In purely mathematical terms it can be obtained as the coadjoint orbit action of the Virasoro group and its extensions \cite{Alekseev:1988ce,Witten:1987ty,Afshar:2019tvp}. Two possible genuine extension/generalization in the gravity setup arises as coupling the 2D dilaton-gravity to new gauge fields and changing the geometry of the spacetime to flat. In this work we are interested in flat space holography and find an extension of the Schwarzian action in the presence of spin-2 gauge fields in the bulk. This provides a first example of higher spin generalization of 2D dilaton-gravity in flat spacetime --- studied in \cite{Afshar:2019axx}, and thus, a first step towards a potential flat-space  higher-spin generalization of the SYK-model.

To this end we study generalizations of the Cangemi-Jackiw model \cite{Cangemi:1992bj} in the BF formulation as a gauge theory of the centrally extended Poincar\'e algebra. Flat space holography in this language is understood as the contraction $\ell\to\infty$ of the AdS algebra with $\ell$ being the AdS radius. However, instead of strictly sending the contraction parameter  $\sigma=\ell^{-1}$ to zero, one may alternatively consider an expansion of the AdS algebra in $1/\ell$. This would extend the Poincar\'e algebra with new generators and new gauge fields appear in the multiplet in each order of the expansion parameter; this is our starting point to flat-space holography in the presence of extra massless interacting gauge fields. In this case the extension would add new spin-1 and spin-2 generators. 

Interacting massless spin-2 gauge fields (or any other massless higher spin interactions with $s\geq2$) coupled to gravity in flat spacetime is forbidden due to no-go theorems. These no-go theorems are however silent in cases with no propagating degrees of freedom like in BF-theories. In general, higher spin theories provide an opportunity to understand holography beyond the supergravity limit which translates to week coupling in the would be SYK-like model.

\subsection*{Summary of results}
The Callan-Giddings-Harvey-Strominger (CGHS) dilaton-gravity model \cite{Callan:1992rs} is an example which admits asymptotically  two-dimensional black holes (for reviews see \cite{Strominger:1994tn,Grumiller:2002nm,Nojiri:2000ja}). The gravity sector of the CGHS model is described by the action
\begin{align}
    I_{\text{\tiny CGHS}}=-\frac{\kappa}{2}\int\extd^2x\sqrt{- g}\,(X R+(\nabla X)^2/X-2X\Lambda )\,.
\end{align} 
By performing a dilaton dependent Weyl rescaling $g_{\mu\nu}\to X^{-1} g_{\mu\nu}$ one obtains the 
 CGHS model in the Einstein frame which is sometime referred to as flat-spacetime JT gravity \cite{Dubovsky:2017cnj,Fitkevich:2020okl}
 \begin{align}\label{FlatJT}
        I_{\text{\tiny CGHS}}\to-\frac{\kappa}{2}\int \extd^2x\sqrt{-g}\,\big(XR-2\Lambda\big)\,,
\end{align}
in which the scalar $X$  enforces $R=0$ as a constraint regardless of the value of $\Lambda$. It can be considered as an effective theory describing the near-horizon properties of non-extremal or near-extremal horizons.  This can be scrutinized
by spherically symmetric reduction of specific 4D gravity models and performing a near-horizon approximation.\footnote{For example, the 4D Einstein-Hilbert term after reduction on sphere with $\extd s_4^2=\extd s_2^2+e^{-2\phi}\extd\Omega_2^2
$ in which the dilaton field is playing the role of the 4D radial direction and making the expansion around a given horizon $r_h=e^{-\phi_h}$ i.e. $e^{-\phi}\sim e^{-\phi_h}(1+X)$,
 leads to \eqref{FlatJT} as a  sub-leading term with $\Lambda=-1/(2r_h^2)$ 
 \begin{align}
        S_{\text{\tiny EH}}\to S_0+\frac{r_h^2}{4\kappa_{4}}\int \extd^2x\sqrt{-g}\,\big(XR+\frac{1}{r_h^{2}}\big)+{\mathcal O}(X^2)\,.
 \end{align}
}

In order to establish a model for flat space holography, in \cite{Afshar:2019axx} following \cite{Cangemi:1992bj} the action \eqref{FlatJT} was marginally manipulated by integrating in a two-dimensional $U(1)$ gauge field $A$ which can be interpreted as a constant electric field and an auxiliary scalar field $Y$. The consequent model was denoted as $\widehat{\text{CGHS}}$ which is equivalent to \eqref{FlatJT} on-shell; \begin{align}\label{CGHShat0}
   I_{\widehat{\text{\tiny CGHS}}}=-\frac{\kappa}{2}\int \extd^2x\sqrt{-g}\big( XR-2Y\big)+\kappa\int Y\extd A\,.
 \end{align}
 The advantage of working with the $\widehat{\text{CGHS}}$ model is that it possesses a BF-theory formulation based on the centrally extended 2D Poincar\'e algebra. More importantly, on-shell the bulk term in \eqref{CGHShat0} is simply zero due to constraints and the full theory is identical to the boundary term. This is reminiscent to the JT gravity while this is not the case in \eqref{FlatJT}.

 The holographic analysis for \eqref{CGHShat0} was performed in \cite{Afshar:2019axx}. It was shown that the theory has a symmetry realization in terms of the twisted warped-conformal algebra \cite{Detournay:2012pc,Afshar:2015wjm} with non-zero twist term but vanishing $U(1)$ level and zero Virasoro central charge. The corresponding Euclidean boundary action was also obtained which coincides with the warped Schwarzian theory \cite{Afshar:2019tvp} which is the geometric action of the same group of centrally extended symmetries namely Diff$(S^1)\ltimes C^\infty(S^1)$ when the Heisenberg subgroup is abelian (vanishing $U(1)$ level) and the Schwarzian term is absent. 
One of the results of our present work is to consider a deformation of the $\widehat{\text{CGHS}}$ model which we call it {\it twisted}-$\widehat{\text{CGHS}}$ model,
\begin{align}\label{CGHShat1}
    I_{\text{\tiny tw-}\widehat{\text{\tiny CGHS}}}=I_{\widehat{\text{\tiny CGHS}}} -\gamma_0\frac{\kappa}{2}\int \extd^2x\sqrt{-g}\,Y R \,,
\end{align}
 which
 is equivalent to the $\widehat{\text{CGHS}}$ model on-shell and yields the Euclidean boundary action as a warped-Schwarzian theory at zero  $U(1)$  level but with non-zero Schwarzian term,
\begin{align}\label{I1eucl}
I_{(1)}^{\text{\tiny E}}=\kappa  \oint \Big( s_0h'^2+\gamma_0\, \text{Sch}(h)+g'\big(i\mathcal P_0 h'-\frac{h''}{h'}\big)\Big)\,,
\end{align}
where $h\in\text{Diff}(S^1)$ is the $S^1$ reparametrization field and $g\in C^\infty(S^1)$ is another field on $S^1$ and 
\begin{align}
    \text{Sch}(h)=\big(\frac{h''}{h'}\big)'-\frac12\big(\frac{h''}{h'}\big)^2\,,
\end{align}denotes the Schwarzian derivative.  The constants $(s_0, \gamma_0,\mathcal P_0)$ specify the corresponding coadjoint orbit of the warped Virasoro group at zero $U(1)$ level.

The main result of this paper, is to extend the construction of $\widehat{\text{CGHS}}$ model in \cite{Afshar:2019axx} by including new spin-2 gauge fields that interact with gravity in flat spacetime. The form of the extended action with only one spin-2 gauge field is,
\begin{align}
\label{L2action0}
 I_{\text{\tiny ex-}\widehat{\text{\tiny CGHS}}}&=
-\frac{\kappa}{2}\int \extd^2x\sqrt{-g}\,\big(X R-2Yf+ 2Y_{(1)}( \nabla_\alpha\nabla_{\beta}
     f^{\alpha\beta}-\nabla^2f-1)
     \big)+\kappa\int Y\extd A\,,
 \end{align}
where a new scalar $Y_{(1)}$ together with the new spin-2 field $f_{\mu\nu}$ are introduced.
In order to obtain \eqref{L2action0}, we extend the Poincar\'e algebra by new generators such that we have a consistently closed algebra that accommodates for the extra spin-2 gauge field and also acquires a well-defined bilinear form so that we can use the gauge theoretic formulation of the non-abelian BF-theory. Again we construct the corresponding geometric action as the Euclidean boundary action of \eqref{L2action0} which turns out to be our main result;
\begin{align}\label{I2eucl}
      I^{\text{\tiny E}}_{(2)}&=\kappa\oint \Big(s_0h'^2+\gamma_0\text{Sch}(h)+g'\big(i{\mathcal P}_0h'-\frac{h''}{h'}\big)+\frac{w'}{h'}\big(t_0h'^2-\gamma_1\text{Sch}(h)\big)+\tfrac12 \big[\big(\frac{w'}{h'}\big)'\big]^2\Big)\,,
\end{align}
where in addition to former fields $h\in \text{Diff}(S^1)$ and $g\in C^{\infty}(S^1)$ here we have an extra field $w\in \text{Vec}(S^1)$ and the orbit is represented by five constants $(s_0,t_0,\mathcal P_0,\gamma_0,\gamma_1)$. The presence of the Schwarzian terms multiplied by $\gamma_0$ and $\gamma_1$ is again indebted to the presence of arbitrary twist terms in the model which we have dropped in \eqref{L2action0} and will be discussed in section \ref{sec3}. 
The geometric action \eqref{I2eucl} is an extension of the former  \eqref{I1eucl} and is associated to the central extension of the extended warped symmetry group;  
\begin{align}\label{extwalgeb}
\text{Diff}(S^1)\ltimes C^\infty(S^1)\ltimes\text{Vec}(S^1)\,,
\end{align}
again with abelian Heisenberg subgroup. It turns out that the appearance of the abelian subgroup is special to flat space holography see e.g. \cite{Barnich:2006av,Barnich:2009se,Bagchi:2010eg,Barnich:2011mi,Barnich:2013axa,Oblak:2016eij,Afshar:2015wjm,Afshar:2013vka,Safari:2019zmc,Parsa:2018kys,Safari:2020pje}. This interesting feature is very different from the generic case of AdS holography in which the corresponding level is non-zero. We exhibit the corresponding centrally extended algebra for the group \eqref{extwalgeb}
 which can be thought of as the spin-2 extension of the warped-conformal algebra at level-zero,
 \begin{align}
  &[L_n,L_m]=(n-m)L_{n+m}+\gamma_0n^3\delta_{n+m,0}\,,\nn\\\vspace{9mm}
  &[L_n,T_m]=(n-m)T_{n+m}+\gamma_1n^3\delta_{n+m,0}\,,\nn\\\vspace{9mm}
  &[L_n,P_m]=-mP_{n+m}+ i\kappa \,n^2\delta_{m+n,0}\,,\qquad
  [T_n,P_m]=0=[P_{m},P_{n}]\,,\nn\\
 &[T_n,T_m]=(n-m)\sum_q P_{m+n-q}P_q+2i\kappa\, (m+n)P_{m+n}+2\kappa^2n^3\delta_{n+m,0}\,.
  \end{align}

\subsection*{Outline}
The structure of this paper is as follows,
In section \ref{sec2}, we use the BF-formulation of dilaton-gravity in 2D flat space and introduce a systematic extension of it with an arbitrary number of interacting spin-2 gauge fields based on the extension of the 2d Poincar\'e algebra.
In section \ref{sec3}, we pay attention to the case with only one spin-2 extension and derive the metric formulation, and discuss the corresponding field equations.
In section \ref{sec4}, we initiate our boundary analysis by discussing the asymptotic symmetries.
In section \ref{varp},  the appropriate boundary terms and integrability conditions for having a well-posed variational principle are found.
In section \ref{sec5}, we switch to the Euclidean signature and find the coadjoint action on the circle.
In section \ref{sec7}, we discuss the thermodynamics of the zero-mode solution both by calculating the Euclidean on-shell action and by implementing the first law using canonical charges.
We summarize and conclude in section \ref{sec8}. In appendix \ref{Appendix1} some aspects of including an infinite number of spin-2 gauge fields in flat space are considered and in appendix \ref{centerapp}, the center of the extended Poincar\'e group is found.

\section{BF formulation of dilaton-gravity in flat space}\label{sec2}
The first-order formulation of a large class of two-dimensional dilaton-gravity theories  has a gauge theory formulation in terms of the \emph{BF-theory}; 
\begin{align}\label{BFtheoryaction}
    I=\kappa\int\left\langle \mathcal B\,,\extd \mathcal A+\tfrac{1}{2}[\mathcal A,\mathcal A]\right\rangle\,,
\end{align}
where $\kappa$ is the coupling constant and the 1-form gauge field $\mathcal A$ and the scalar $\mathcal B$ are Lie algebra valued see e.g. \cite{Fukuyama:1985gg,Isler:1989hq,Schaller:1994es}. The pairing between $\mathcal B$ and the curvature 2-form is via a non-degenerate bilinear form of the Lie algebra. The field equations demand the curvature two form to be zero; so no local degrees of freedom are involved. 
As in all dimensions, two-dimensional gravity multiplet consists of the spin-connection $\omega$ and the zweibein $e^a$ one-forms so the connection is,
\begin{align}
    \mathcal A=e^aP_a+\omega J+\cdots\,,
\end{align}
where $J$ and $P$'s are the boost and translation generators in 2D. The $\cdots$ are possible extra gauge fields. Unlike higher dimensions, the gravity multiplet needs to be supplemented with the dilaton multiplet which is taken into account by the $\mathcal{B}$ field.

The Jackiw-Teitelboim gravity in its first order formulation is equivalent to a BF-theory with the gauge algebra $\mathfrak{so}$(2,1) with no extra gauge symmetry.\footnote{For BF-theory based on higher-spin gauge symmetry in JT gravity see \cite{Fradkin:1989uh,Bengtsson:1986zm,Vasiliev:1995dn,Alkalaev:2013fsa,Grumiller:2013swa,Gonzalez:2018enk,Alkalaev:2019xuv,Alkalaev:2020kut}.}
In flat spacetime a similar construction allowed Cangemi and Jackiw to write a BF-theory \cite{Cangemi:1992bj} based on the  central extension of the 2D Poincar\'e algebra $\mathfrak{iso}$(1,1) namely Maxwell algebra 
\begin{align}\label{maxwell2}
[P_a,J]=\epsilon_a{}^bP_b\,,\qquad[P_a,P_b]=\epsilon_{ab}J^{(1)}\,,
\end{align}
with  $J^{(1)}$ appearing as a central extension. The crucial fact about this algebra unlike its non-central cousin is the appearance of a non-degenerate bilinear form which enables us to write the BF-theory. Unlike JT-gravity where the multiplet contains only a graviton and a dilaton, here in flat spacetime, the multiplet contains a graviton, a dilaton, and a single spin-1 gauge field whose total degrees of freedom sum to zero.  The presence of the extra spin-1 is a consequence of the extra generator introduced as the central extension of the algebra.

If we intend to add a new spin-2 gauge field to the BF-system we need to enlarge the algebra such that new generators transform covariantly under the boost generator $J$,
\begin{align}
[P_a^{(1)},J]=\epsilon_a{}^bP^{(1)}_b\,.
\end{align}
This is obviously a necessary condition but not sufficient. In principle, we need to introduce as many other generators so that we can close the algebra and simultaneously make sure that a new non-degenerate invariant bilinear form emerges. Although in principle this strategy works, here we develop a systematic approach based on the extension of the Poincar\'e algebra.

\subsection{Extension of the Poincar\'e algebra}
Here, we focus on the extension of the two-dimensional Poincar\'e algebra but our strategy is more general and can be applied to any dimensions. Moreover, we turn attention to the relativistic case while non-relativistic algebras can be discussed on an equal footing.\footnote{For a non-relativistic 2d dilaton-gravity setup see the recent papers \cite{Grumiller:2020elf,Gomis:2020wxp,Hansen:2020hrs}.} 
We use the fact that the two dimensional Poincar\'e algebra is a contraction of the  $\mathfrak{so}(2,1)$ algebra which has three generators $P_0$, $P_1$  and $J$  forming the gauge algebra for (A)dS$_2$ JT-gravity,
\begin{align}\label{so12}
    [P_a,J]=\epsilon_a{}^bP_b\,,\qquad[P_a,P_b]=  \epsilon_{ab}J\,,
\end{align}
where we have fixed the cosmological constant to $-1$.
One can introduce three isomorphic contractions of this algebra by appropriately rescaling only one of these generators with a contraction parameter $\sigma$ and then send it to  zero in the algebra. This leads to three distinct but isomorphic algebras namely 2D Poincar\'e (relativistic), 2D Newton-Hooke (non-relativistic) and Carroll AdS$_2$ (ultra-relativistic). 

Alternatively, one can preserve the contraction parameter in this process and use it as an expansion parameter rather than just approaching zero. In this procedure all new generators appearing at each order in the contraction parameter are independent. This procedure leads to an extension of the contracted algebra. In the relativistic case we should introduce the contraction parameter $\sigma$ in the last commutator of \eqref{so12} which is viewed now as an In\"on\"u-Wigner (IW) bundle of the Lie algebra $\mathfrak{so}(2,1)$ over an affine line parametrized by $\sigma$ \cite{Khasanov:2011jr}.
For this IW bundle we have $J^{(0)}\equiv J$, $J^{(n)}\equiv\sigma^n J$  and $P_{a}^{(n)}\equiv\sigma^n P_{a}$.
In general, this system introduces an arbitrary extension of the contracted algebra with the following commutators
\begin{align}\label{infinite-poinccare}
    [P^{(m)}_a,J^{(n)}]=\epsilon_a{}^bP^{(m+n)}_b\,,\qquad[P^{(m)}_a,P^{(n)}_b]= \epsilon_{ab}J^{(m+n+1)}\,.
\end{align}
The level zero is  the Poincar\'e algebra
\begin{equation}
    [P_a,J]=\epsilon_a{}^b P_b\,,\qquad [P_a,P_b]=0\,.
\end{equation}
However at this level, $P_a$ generators form an ideal for the algebra so we miss an invariant bilinear form. This problem is resolved at level-1 where the algebra acquires a new generator $J^{(1)}$ such that we are led to the 2D Maxwell algebra \eqref{maxwell2}. This algebra admits an invariant non-degenerate bilinear form 
\begin{align}\label{g1}
    \langle J,J^{(1)}\rangle=1\,,\qquad \langle P_a,P_b\rangle=-\eta_{ab}\,,
\end{align}
The BF-theory based on this algebra and the bilinear form \eqref{g1} was introduced in \cite{Cangemi:1992bj,Verlinde:1991rf}. One can deform the model using the fact that the bilinear form at this level is more general than \eqref{g1} and a new invariant can be added to it \cite{Nappi:1993ie}
\begin{align}\label{JJ}
   \langle J,J\rangle=\gamma_0\,.
\end{align}
The BF-theory based on this more general metric will be discussed in section \ref{sec3} and the corresponding boundary dynamics is addressed in section \ref{sec5}. 

At level-2, we can add three new generators as $J^{(2)}$, $P^{(1)}_a$ to \eqref{maxwell2} which satisfy  new non-vanishing kinematical commutators 
\begin{align}\label{level2alg1}
 [P_{a}^{(1)},J]=\epsilon_a{}^bP_{b}^{(1)}\,,\qquad
[P_{a},J^{(1)}]=\epsilon_a{}^bP_{b}^{(1)}\,,\qquad[P_{a},P_{b}^{(1)}]=\epsilon_{ab}J^{(2)}.
\end{align}
We can see that at this level $J^{(1)}$ is not a central term any more and the new generator $J^{(2)}$ is central. The algebra at this level in general possesses the invariant bilinear forms of the level-0 \eqref{JJ} and the level-1 \eqref{g1} which by themselves are degenerate at this level but together with the following new bilinear form they form an invariant non-degenerate bilinear form at level-2
\begin{align}\label{innerprod}
   \langle J,J^{(2)}\rangle=1 \,,\qquad \langle\, P_a,P^{(1)}_b\rangle=-\eta_{ab}\,,\qquad \langle J^{(1)},J^{(1)}\rangle=1\,.
\end{align}
In principle, one can continue this procedure and extend the algebra to higher levels 
as will be discussed below.
\subsection{Casimir and the bilinear form}\label{App2}
In principle, one can continue the procedure of generating new generators to the level-$N$. Denote by $J^{(N)}$ and $P_a^{(N-1)}$ as the generators included in the $N$-th  extension of the Poincar\'e algebra. The algebra at this level contains $3N+1$ generators and we show it by $\mathfrak{g}^{(N)}$.  Suppose that the generators for a given level $N$ have the following order
\begin{equation}
    J,\, P_a,\, J^{(1)},\,P_a^{(1)},\, J^{(2)},\, \cdots\,,P_a^{(N-1)},\, J^{(N)}\,.
\end{equation}
In particular, we have a right shift on this sequence by commuting with $P_a$,  until it hits $J^{(N)}$ where the sequence ends and gives zero.

 The algebra at level $N$ possesses a linear Casimir $C_0=J^{(N)}$ as well as $N$ quadratic Casimir operators. In the latter set, the one that includes all generators is the following
 \begin{equation}
     C_{N}=\frac{1}{2}\sum_{k=0}^N\big(J^{(k)}J^{(N-k)}-\eta^{ab}P_a^{(k)}P_b^{(N-k-1)}\big)\,.
 \end{equation}
 The other $N-1$ quadratic Casimirs are
 \begin{equation}
     C_{I}=\frac{1}{2}\sum_{k=0}^N\big(J^{(k+N-I)}J^{(N-k)}-\eta^{ab}P_a^{(k+N-I)}P_b^{(N-k-1)}\big)\,,\qquad I=1,\cdots,N-1\,.
 \end{equation}
For the case $N=1$ the Maxwell algebra has one linear $C_0=J^{(1)}$ and one quadratic Casimir
\begin{align}
     C_{1}&=JJ^{(1)}-\frac{1}{2}P^aP_a\,.
\end{align}
At level $N=2$, we have a linear $C_0=J^{(2)}$ and two quadratic Casimirs
\begin{align}\label{casimir222}
{C_{1}}&=J^{(2)}J^{(1)}-\frac{1}{2}P^a P_a\,,\nn\\
{C_{2}}&=JJ^{(2)}+\frac{1}{2}J^{(1)}J^{(1)}-P^a P^{(1)}_a\,.
\end{align}
There is a one to one map between these Casimirs and the invariant bilinear forms of the algebra at each level. Namely, the Casimir is the inverse of the bilinear form up to a constant since it is mapped to a constant under the action of the bilinear form
\begin{align}
     \Gamma: \mathfrak{g}^\ast\times \mathfrak{g}\to\mathbb{R}\,.
\end{align}
Therefore, at level $N$, we have $N+1$ independent bilinear forms, $\{\Gamma_0\cdots\Gamma_N\}$, such that $\Gamma_k$ is the inverse of $C_{k}$ for $k=1,\cdots,N$ and $\Gamma_0$ is the inverse of $(J^{(N)})^2$. Among thses, only $\Gamma_N$ is independently non-degenerate. For example, we have two independent bilinear forms at level one
\begin{equation}\label{level1 bilinear}
\begin{array}{ll}
    \Gamma_0:\qquad\qquad & \langle J,J\rangle={\gamma_0}
\\
     \Gamma_1:\qquad\qquad&  \langle J,J^{(1)}\rangle={\gamma_1}\qquad\langle P_a,P_b\rangle=-{\gamma_1}\eta_{ab}\,.
  \end{array}
   \end{equation}
and  three invariant bilinear forms at level-2 
\begin{equation}\label{bilinear2}
\begin{array}{lll}
    \Gamma_0:\qquad\qquad & \langle J,J\rangle={\gamma_0}&
\\
     \Gamma_1:\qquad\qquad&  \langle J,J^{(1)}\rangle={\gamma_1}&\qquad\langle P_a,P_b\rangle=-{\gamma_1}\eta_{ab}\\
      \Gamma_2:\qquad\qquad&    \langle J^{(1)},J^{(1)}\rangle={\gamma_2}&\qquad\langle P_a,P^{(1)}_b\rangle=-{\gamma_2}\eta_{ab}
    \qquad\langle J,J^{(2)}\rangle={\gamma_2}\,.
  \end{array}
   \end{equation}
In our construction of BF-theories, we use the most general bilinear map, labeled by the real $N$-vector $\boldsymbol{\gamma}=({\gamma_0,\gamma_1,\cdots\gamma_{N-1}})$, such that
\begin{equation}\label{genbilin}
    \Gamma=\Gamma_N+\sum_{k=0}^{N-1}\gamma_k\Gamma_k\,.
\end{equation}
The coefficient $\gamma_N$ behind $\Gamma_N$ must be non-zero and is set to one, while $\gamma_k$ are arbitrary real numbers.
In \eqref{genbilin} the bilinear form $\Gamma_k$ is
\begin{equation}\label{Nbilin}
 \langle P_a^{(n)},P_b^{(m)}\rangle=-\eta_{ab}\delta_{n+m-k}\,,\qquad\qquad    \langle J^{(n)},J^{(m)}\rangle=\delta_{n+m-k}\,,
\end{equation}
where $k\leq N$.

\subsection{Extended BF-theory formulation}
At the moment we have provided a consistent extension of the 2D Poincar\'e algebra.
In the frame-like formulation of the extended dilaton-gravity we should complement the gravity multiplet, the zweibein 1-form $e$ and the spin connection 1-form $\omega$ with new one-forms which play a similar role for the extra spin-2 gauge symmetries.  In general for the extended gauge group discussed above we can consider the following  gauge connection at level-$N>0$,
\begin{align}\label{BFgaugen}
    \mathcal A_{(N)}=\mathbf{e}+\boldsymbol{\omega}+ A J^{(N)}\,.
\end{align}The corresponding field strength for the gauge field \eqref{BFgaugen} can be written as \begin{align}
    \mathcal F_{(N)}=T(\mathbf{e})+R(\boldsymbol{\omega})+F J^{(N)}\,,
\end{align}
In addition to the gauge field \eqref{BFgaugen} we also introduce a scalar field at level-$N>0$,
\begin{align}\label{BFscalrn}
    \mathcal B_{(N)}=\mathbf{Z}+\mathbf{Y}+ X J^{(N)}\,.
\end{align}
In \eqref{BFgaugen} and \eqref{BFscalrn} the $U(1)$ gauge field $A$ and the scalar field $X$ corresponding to the central term of the algebra $J^{(N)}$ are distinguished as they are independent fields. In fact, the scalar field $X$ in the second-order form will play the role of the dilaton field coupled to gravity. Since $J^{(N)}$ appears as the central term in the commutator of $P_a$ and $P_a^{(N-1)}$, the gauge field $A$  is like a St\"ueckelberg field that shift-transforms under the gauge transformation along  $P_a^{(N-1)}$ to restore gauge invariance: setting $A$ to zero is inconsistent with gauge symmetry of the theory.
For extra components in \eqref{BFgaugen} and \eqref{BFscalrn} we have
\begin{subequations}\label{bfgaug2}
\begin{align}
    \mathbf{e}&=e^a P_a+e^a_{(1)}P_a^{(1)}+\cdots+e^a_{(N-1)}P_a^{(N-1)}\,,\\
    \boldsymbol{\omega}&=\omega J+ \omega_{(1)}J^{(1)}+\cdots+\omega_{(N-1)}J^{(N-1)}\,,
\end{align}
\end{subequations}
and 
\begin{subequations}\label{bfdil}
\begin{align}
    \mathbf{Z}&=Z^a P_a+Z^a_{(1)}P_a^{(1)}+\cdots+Z^a_{(N-1)}P_a^{(N-1)}\,,\\
    \mathbf{Y}&=Y J+ Y_{(1)}J^{(1)}+\cdots+Y_{(N-1)}J^{(N-1)}\,.
\end{align}
\end{subequations}
In the expression for the gauge field \eqref{bfgaug2} the zweibein $e$ and extra $e_{(k)}$ are independent gauge fields while $\omega$ as the spin-connection and extra $\omega_{(k)}$ are dependent ones. Their dependence is fixed via torsion constraints $T(\mathbf{e})=0$ which can be integrated in the action via the Lagrange multipliers $\mathbf Z$. Finally, the scalar fields in $\mathbf{Y}$ will play the role of auxiliary scalar fields which couple to $R(\boldsymbol{\omega})$ and play the role of extended dilatons. This construction can be written as a BF-action. In fact, as was shown in subsection \ref{App2} our extension of the 2D Poincar\'e algebra is such that at each level of the extension we have a non-degenerate bilinear form. This suggests that we can consistently extend the corresponding BF-theory.
At level-$N$, all the generators associated with the extended algebra $\mathfrak{g}^{(N)}$ contribute, and the action is formally given by
\begin{align}\label{BFseries}
 I_{(N)}= \gamma_0\text{BF}^{(0)}+\gamma_1\text{BF}^{(1)}+\cdots+\gamma_{N-1}\text{BF}^{(N-1)}+\text{BF}^{(N)}\,.
\end{align}
where we have extracted the arbitrary coefficients $\gamma_k$. The $\text{BF}^{(k)}$ in the above sequence is defined by using the bilinear form $\Gamma_k$ in \eqref{genbilin}
\begin{align}
    \text{BF}^{(k)}=\int \left\langle{\mathcal B}_{(N)},{\mathcal F}_{(N)}\right\rangle_{\Gamma_k}\,.
\end{align}
Note that at level $N$ we have the right to set to zero or keep as many of $\gamma_{k}$ for $k<N$. For example, at level-$1$ we have an arbitrary parameter $\gamma_0$ and at level-$2$ we have two unfixed parameters $\gamma_0$ and $\gamma_1$, and this pattern persists in higher levels. 
Although all these choices are equivalent on-shell,  they correspond to different theories off-shell, and especially their boundary dynamics are different. This will be shown later in two examples of $N=1,2$ in section \ref{sec3}.  
We ponder the case of $N\to\infty$ in appendix \ref{Appendix1}.

\section{Extended dilaton-gravity in flat space}\label{sec3}
In this section, we use the extension of the 2D Poincar\'e algebra discussed above to construct the dilaton-gravity at level-2 which exhibits an extra spin-2 gauge symmetry. 
The construction at level-1 is reviewed in advance.

At level-1 we deal with the algebra given in \eqref{maxwell2} and the bilinear form is introduced in equations \eqref{g1}-\eqref{JJ}. The gauge field and the scalar field at this level can be read from eq. \eqref{BFgaugen} and \eqref{BFscalrn}. We can then systematically write down the BF-theory for the 2D Maxwell algebra as the first order formulation of dilaton-gravity in flat space at level-1
    \begin{align}\label{BFseries1}
 \kappa^{-1}I_{(1)}= \gamma_0\int Y\extd \omega+\int \big[X\extd \omega+Y(\extd A+ \tfrac{1}{2}\epsilon_{ab}\,e^ae^b)-Z_a(\extd e^a+\epsilon^a{}_b\,\omega\, e^b)\big]\,.
\end{align}
The second order action corresponding to this model is simply obtained by solving the torsion constraint {and using the fact that in two dimensions $\extd\omega=-\frac12 R\epsilon$ with $\epsilon$ being the volume form,}
\begin{align}\label{CGHShat}
   \kappa^{-1} I_{(1)}= -\frac{\gamma_0}{2}\int \extd^2x\sqrt{-g}\,Y R-\frac{1}{2}\int \extd^2x\sqrt{-g}\big[ XR-2Y\big]+\int Y\extd A\,.
\end{align}
The case where $\gamma_0=0$ corresponds to the $\widehat{\text{CGHS}}$ model which was 
studied from the holographic point of view in \cite{Afshar:2019axx}\footnote{{The overall sign-difference of the action here and in    \cite{Afshar:2019axx} is due to our convention of bilinear form \eqref{level1 bilinear}.}}. We denote the case $\gamma_0\neq0$ as the {\it twisted}-$\widehat{\text{CGHS}}$ model whose boundary dynamics will be discussed later. The field equations for both models are the same and are given in \cite{Afshar:2019axx} 
\begin{align}
    R=0\,,\qquad
    \varepsilon^{\mu\nu}\partial_\mu A_\nu=1\,,\qquad
    \nabla_\mu\nabla_\nu X+g_{\mu\nu}Y=0\,,\qquad Y=\text{const.}
\end{align}
where appropriate solutions are also discussed. Here $\varepsilon^{\mu\nu}=\frac{\epsilon^{\mu\nu}}{\sqrt{-g}}$ is the $\varepsilon$-tensor.
\subsection{Extended-\texorpdfstring{$\widehat{\text{CGHS}}$}{CGHS} model}\label{ExtendedCGHS}
Here we consider the extension of the 2D Poincar\'e algebra at level-2 with commutators given in \eqref{maxwell2} and \eqref{level2alg1} and the  bilinear form introduced in section \ref{sec2} in equation \eqref{bilinear2}. One can systematically construct the  BF-formulation of the extended dilaton gravity at this level using the gauge field \eqref{BFgaugen}  and the scalar field \eqref{BFscalrn}
\begin{align}\label{actionlevel2}
   \kappa^{-1} I_{(2)}&=\gamma_0\int Y\extd \omega+\gamma_1\int \big[Y_{(1)}\extd \omega+YR_{(1)}-Z_aT^a\big]\nn\\&+\int \big[X\extd \omega+YF+Y_{(1)}  R_{(1)}-Z^a_{(1)} T_a-Z_aT^a_{(1)} \big]\,,
\end{align}
where the curvature 2-forms are introduced  as follows
\begin{subequations}
\begin{align}
F&=\extd  A+ \epsilon_{ab}\,e^a e_{(1)}^b\,,\\
     R_{(1)}&=\extd \omega_{(1)}+ \tfrac{1}{2}\epsilon_{ab}\,e^ae^b\,,\\
    T^a&=\extd e^a+\epsilon^a{}_b\,\omega\, e^b\,,\\
    T^a_{(1)}&=\extd e_{(1)}^a+\epsilon^a{}_b\,\omega\, e_{(1)}^b+\epsilon^a{}_b\,\omega_{(1)} e^b
    \,.
\end{align}
\end{subequations}
In order to migrate to the metric formulation we integrate out in \eqref{actionlevel2} the Lagrange multipliers $Z^a$ and $Z^a_{(1)}$ which gives $T^a=T_{(1)}^a=0$, thereby the dependent gauge fields $\omega$ and $\omega_{(1)}$ can be obtained,
    \begin{subequations}\label{Tor}
    \begin{align}
    \omega&=e_a\star\extd e^a=\frac{1}{2}e_a(\extd e^a)_{\sigma\rho}\varepsilon^{\sigma\rho}\\
     \omega_{(1)}&=e_a\star\mathcal  De_{(1)}^a=\frac{1}{2}e_a(\mathcal De_{(1)}^a)_{\sigma\rho}\varepsilon^{\sigma\rho}\,,
\end{align}
    \end{subequations}
    where $\mathcal D=\extd +[\omega,\,]$ is the Lorentz  covariant exterior derivative. 
    Here  we used the fact that in two dimensions every antisymmetric tensor is proportional to $\varepsilon$. These solutions transform appropriately under a general gauge transformation
\begin{align}\label{generalgaugetrans}
\Lambda=
\laa J+\xi^aP_{a}+\laa_{(1)}J^{(1)}+\xi_{(1)}^aP_a^{(1)}+\sigma J^{(2)}\,,
\end{align}
Once we use the gauge transformation of independent gauge fields as 
\begin{subequations}\label{gaugetrans1}
\begin{align}
    \delta e^a&=
    \mathcal{D}\xi^a-{\epsilon^a}_b\laa e^b\,,\\
       \delta e_{(1)}^a&=\mathcal{D}\xi_{(1)}^a+{\epsilon^a}_b \omega_{(1)}\xi^b-{\epsilon^a}_b\laa e_{(1)}^b-{\epsilon^a}_b\laa_{(1)} e^b
              \,,\\
             \delta A&=\extd\sigma -\epsilon_{ab}e_{(1)}^a\xi^b+\epsilon_{ab}e^a\xi_{(1)}^b
       \,.
\end{align}
\end{subequations}
We express the metric $g_{\mu\nu}$ and the spin-2 gauge field $f_{\mu\nu}$ in transition to the second order formulation  such that they are invariant under the Lorentz boost $\lambda$ and the extended Lorentz boost $\lambda^{(1)}$ appearing in \eqref{gaugetrans1},\footnote{We use the symmetrization convention as $X_{(\mu\nu)}=X_{\mu\nu}+X_{\nu\mu}$.}
\begin{align}\label{metricspin2}
    g_{\mu\nu}=\frac{1}{2} e_{(\mu}\cdot e_{\nu)}\,,\qquad\qquad f_{\mu\nu}=\frac{1}{2} e_{(\mu}\cdot e^{(1)}_{\nu)}\,,
\end{align}
{where $\cdot$ represents contraction of flat indices with $\eta_{ab}$. Here, we base our second order theory by identifying its metric and the extra spin-2 gauge field as in \eqref{metricspin2}. Does there exists other consistent theories? To answer this question one could try to find other consistent embeddings of the Poincar\'e iso(1,1) algebra into the extended Poincar\'e algebra given in \eqref{maxwell2} and \eqref{level2alg1} and its corresponding bilinear form \eqref{bilinear2}. Here we considered the principle embedding where $P_a$'s play the role of spacetime translation. There exists two other non-principle choices where ($P_0$, $P_1^{(1)}$) or ($P_0^{(1)}$, $P_1$) can be identified with spacetime translation. It would be interesting to explore these choices separately.}
The expression in \eqref{Tor} and \eqref{metricspin2} enables us to write the exterior derivatives of the spin-connection $\omega$ and the extended connection $\omega_{(1)}$ appearing in  \eqref{actionlevel2}  in terms of the  two dimensional Ricci scalar $R$ and covariant derivative of the spin-2 gauge field $f_{\mu\nu}$ and its trace $f$,
\begin{subequations}\label{extdomega}
\begin{align}
    \epsilon^{ab}\extd\omega&=\frac{1}{2}R\,e^ae^b\,,\\
    \epsilon^{ab}(\extd\omega_{(1)})_{ab}&=2\left(\mathcal D^2 f-\mathcal D_a\mathcal D_b f^{ab}\right)\,.
\end{align}
\end{subequations}
Inserting \eqref{extdomega} into the action \eqref{actionlevel2}, we have its second order form
 \begin{align}
\label{L2action}
 \kappa^{-1}I_{(2)}&=
      -\frac{\gamma_0}{2}\int \extd^2x\sqrt{-g}\, Y R
 -\frac{\gamma_1}{2}\int \extd^2x\sqrt{-g}\,\big[ Y_{(1)} R
  +2Y( \nabla_\alpha\nabla_{\beta}
     f^{\alpha\beta}-\nabla^2f-1)\big]\nn\\
&\,\quad-\frac{1}{2}\int \extd^2x\sqrt{-g}\,\big[X R+ 2Y_{(1)}( \nabla_\alpha\nabla_{\beta}
     f^{\alpha\beta}-\nabla^2f-1)
     +2Y(\varepsilon^{\alpha\beta}
     \partial_\alpha{A_\beta}-{f})\big]\,.
 \end{align}
This action is obviously invariant under diffeomorphism that act on the fields via the Lie derivative. In addition, there are spin-2, as well as  $U(1)$ gauge symmetries generated by gauge parameters $\chi$ and $\sigma$ correspondingly, which transform the spin-2 gauge field $f_{\mu\nu}$ and the spin-1 gauge field $A_\mu$ and also the dilaton field $X$,\footnote{The variation of the dilaton field $X$ cancels the curvature contribution  induced by commuting the covariant derivatives in variation of $f_{\mu\nu}$ under $\chi_\mu$.}
\begin{align}\label{gaugetr}
    \delta f_{\mu\nu}=\nabla_{(\mu}{\chi}_{\nu)}\,,\qquad
   \delta A_\mu=\partial_\mu\sigma+ 2{\varepsilon^\nu}_\mu\chi_\nu\,, \qquad\delta X=2\chi^\mu\nabla_\mu (Y_{(1)}+\gamma_1Y),
\end{align}
while keeping $g_{\mu\nu}$, $Y_{(1)}$ and $Y$ invariant.  This shows that in order to preserve the spin-2 gauge symmetry off-shell, the dilaton field $X$ should also transform   under the spin-2 gauge transformation $\chi$ as well as the gauge field $A_\mu$ which apart form its $U(1)$ gauge symmetry generated by $\sigma$, shift transforms under $\chi$.

\subsection{Equations of motion}\label{subsec3}
Upon varying the action with respect to the scalars $X$, $Y_{(1)}$ and $Y$ we get  the following constraints on gauge fields 
 \begin{subequations}\label{eom12}
\begin{align}
   R&=0\,,    \\
   \nabla_{\beta}
    \nabla_\alpha f^{\alpha\beta}-\nabla^2f&=1 \,,\\   \varepsilon^{\alpha\beta}
    \partial_\alpha{A_\beta}&={f}\,, 
\end{align}
and upon varying the metric $g_{\mu\nu}$, the spin-2 gauge field $f_{\mu\nu}$ and the spin-1 gauge field $A_\mu$ we get the following equations for scalar fields
\begin{align}
  \nabla_{\mu}\nabla_{\nu}X-g_{\mu\nu}\nabla^2 X&=2Y_{(1)}\big(\nabla_\sigma\nabla_{(\mu} {f_{\nu)}}^\sigma
-\nabla_\mu\nabla_\nu f-\nabla^2f_{\mu\nu}\big)
-2Y\big(f_{\mu\nu}-\frac{1}{2}fg_{\mu\nu}\big)
 \,,\\
    \nabla_{\mu}\nabla_{\nu}Y_{(1)}+g_{\mu\nu}Y&=0\,,\\
Y&=\text{const.}
\end{align}
\end{subequations}
where in the first line above we imposed the first two constraints and in the second line we have subtracted the trace.
We solve now some of the field equations of the extended-$\widehat{\text{CGHS}}$  \eqref{eom12} with specific suitable boundary and gauge fixing conditions. 
\paragraph{Metric.}
{The starting point is to specify the boundary conditions on the metric. We set up our boundary conditions with Rindler behavior which in the outgoing Eddington–Finkelstein gauge ($g_{ur}=-1$ and $g_{rr}=0$) is represented as $\extd s^2\sim \mathcal{O}(r)\extd u^2-2\extd u\extd r$.  The spacetime is  Ricci flat $R=0$, so  we take the general solution  \cite{Afshar:2019tvp}}
\begin{equation}\label{EF metric}
    \extd s^2=-2\extd u\extd r+2(\cT(u)+r\cP(u))\extd u^2\,.
\end{equation}
{As usual, regularity  at the horizon fixes the zero-mode of the leading term in terms of the Rindler acceleration or the horizon temperature. This means that the zero mode of $\mathcal P$ is fixed on the disk. The zero mode of $\mathcal T$ corresponds to the mass of the black holes. The location of the horizon in \eqref{EF metric} is at $r_{\text{\tiny H}}=-\frac{\cT_0}{\cP_0}$.}

We can identify the zweibein components form \eqref{EF metric} in the light-cone frame as
\begin{align}\label{lightconezweibein}
    e^+=(\mathcal P(u)r+\mathcal T(u))\extd u-\extd r\,,\qquad e^-=\extd u\,.
\end{align}
Using \eqref{lightconezweibein} and the torsion constraint $T^a=0$ we simply have the spin-connection as $\omega=\mathcal P\extd u$. 
\paragraph{Spin-2 gauge field.} In order to solve for the field $f_{\mu\nu}$ we  use the axial gauge $f_{ur}=f_{rr}=0$. The equations of motion implies $\partial^2_r f_{uu}=1$ which can be solved as
\begin{equation}\label{fsolution}
 f_{uu}=\cTi(u)+r\cPi(u)+\tfrac{1}{2}r^2.
\end{equation}
The frame components are given as
\begin{align}
   e^+_{(1)}=\big(\tfrac{1}{2}r^2+r\cPi(u)+\cTi(u)\big)\extd u\,,\qquad  e^-_{(1)}=0\,.
\end{align}
The torsion constraint $T^a_{(1)}=0$ implies the corresponding spin-2 connection $\omega_{(1)}=(r+\mathcal P_{(1)})\extd u$. 
\paragraph{Spin-1 gauge field.} The equation of motion for $A$ is equivalent to \begin{align}
    \partial_\mu A_\nu-\partial_\nu A_\mu=-f\epsilon_{\mu\nu}\,,
\end{align}
    where $f$ is the trace of $f_{\mu\nu}$. In the gauge \eqref{fsolution} where $f=0$, the $U(1)$ gauge field is a pure gauge and can be gauge fixed to $A=0$. 
\paragraph{Dilaton fields.} In the axial gauge for $f_{\mu\nu}$ the $rr$-component of the field equation for $Y_{(1)}$ can be easily solved
\begin{align}\label{Ydilaton}
    Y_{(1)}=y_1(u)r+y_0(u)\,.
\end{align}
The $rr$ component of the field equation for $X$ also gives $\partial_r^2X=0$ which is simply solved as;
\begin{equation}\label{Xdilaton}
    X=x_1(u)r+x_0(u)\,.
\end{equation}
We continue our discussion on field equations, the boundary action as well as asymptotic symmetry analysis in the gauge theory BF-formulation in the next section \ref{sec4}.

\section{Boundary analysis}\label{sec4}
In this section, we aim to derive the boundary action upon introducing a consistent set of boundary conditions. This derivation is based on the BF-formulation of the model. In this construction, the boundary action is obtained upon establishing a well-defined variational principle in the BF-model. This translates to making the boundary term on the imposed boundary conditions integrable. 
\subsection{Boundary conditions and symmetries}
Before discussing the variational principle we present the boundary conditions and the corresponding symmetries in the gauge theory BF-formulation.
\paragraph{Light-cone.} In order to proceed we find it useful to go to the light-cone gauge. In this gauge $\eta_{+-}=1$ and $\epsilon_{\pm}{}^{\pm}=\pm1$ so the extended gauge algebra at level-2 can be written as\footnote{The relation to the diagonal gauge is $P_\pm=(P_{{1}}\pm P_{{0}})/\sqrt{2}$ and $P^{(1)}_\pm=(P^{(1)}_{{1}}\pm P^{(1)}_{{0}})/\sqrt{2}$.}
\begin{align}
  [ P_\pm,J]&=\pm P_\pm\qquad
  [ P_+,P_-]= J^{(1)}\label{Maxwel1}
  \\
  [ P_\pm,P^{(1)}_\mp]&= \pm J^{(2)}
  \qquad[ P^{(1)}_\pm,J]=\pm P^{(1)}_\pm\qquad   [ P_\pm,J^{(1)}]=
  \pm P^{(1)}_\pm\,,\label{level2alglc}
\end{align}
and the bilinear form  at level-2 \eqref{bilinear2} takes the form
\begin{align}
    &\langle J,J\rangle=\gamma_0\,,\label{gggggg}\\
    &\langle J,J^{(1)}\rangle=-\langle P_+,P_-\rangle=\gamma_1\,,\label{JJJJJJ}\\
    & \langle J,J^{(2)}\rangle=\langle J^{(1)},J^{(1)}\rangle=-\langle  P_+,P^{(1)}_-\rangle= -\langle  P_-,P^{(1)}_+\rangle=1\,.\label{bilinearL2}
\end{align}
We setup our boundary conditions compatible with the metric formulation in subsection \ref{subsec3} for the metric $g_{\mu\nu}$ in \eqref{EF metric} and for the spin-2 gauge field $f_{\mu\nu}$ in \eqref{fsolution} and the fact that the gauge field $A_\mu$ is pure gauge. After translating those boundary conditions to the first order form, inserting into \eqref{BFgaugen} and 
performing a finite state-independent gauge transformation
\begin{align}
    \mathcal A_\mu\to g\left(\partial_\mu+\mathcal A_\mu\right) g^{-1}\,,\qquad g=\exp(-r P_+)\,,
\end{align}
 the $r$-dependence in the gauge field disappears and we are left with non zero elements,
\begin{align}\label{L2bc}
    \mathcal A_u&=\mathcal P\,J+\mathcal T\,P_{+}+P_-+\mathcal P_{(1)}J^{(1)}+\mathcal T_{(1)}P_+^{(1)}\,.
\end{align}
We choose a gauge fixing condition and justify it in subsection \ref{subsec42},
\begin{align}\label{gaugefixingP1}
    \mathcal P_{(1)}=0\,.
\end{align}
Using the same transformation on the $\mathcal B$ field $\mathcal B\to g\mathcal B \,g^{-1}$ we can decompose it in terms of the generators of the algebra and the dilaton fields in \eqref{Ydilaton} and \eqref{Xdilaton},
\begin{align}\label{Bcompn}
\mathcal B=YJ
+y^+P_{+}+y_1P_-+y_0J^{(1)}
+x^+P_+^{(1)}+x_1P_-^{(1)}+x_0J^{(2)}\,,
\end{align}
{where $x^+$ and $y^+$ are determined by field equations.} All functions in the connection $\mathcal A$ and the $\mathcal B$-field are allowed to vary {and are functions of $u$. Since the Lie algebra is larger at level-2, more functions appear in the gauge field and the dilaton, and there are additional field equations that determine the dynamics.}
On-shell, the scalar field $\mathcal B$ plays the role of the stabilizer for $\mathcal A_u$,
\begin{align}\label{eom}
     \mathcal B' +[\mathcal A_u,\mathcal B]=0\,.
\end{align}
where prime is the derivative w.r.t. $u$. 
These field equations solve four components of the scalar field in terms of other components;\footnote{ Generalization of this boundary condition to level-$N$ where we have $3N+1$ generators, solves $2N$ variables of the $\mathcal B$-field in terms of $N+1$ free variables.}
\begin{subequations}\label{eomB1}
\begin{align}
    y^+&=y_0'+\mathcal T y_1\,,\\
    x^+&=x_0'+\mathcal T_{(1)} y_1+\mathcal T x_1\,,\\
    Y&=y_1'+\mathcal P y_1\,,\\
   y_0&=x_1'+\mathcal P x_1\,,
\end{align}
\end{subequations}
We have two more equations which relates different components to each other;
\begin{subequations}\label{eomB2}
\begin{align}
    &(y^+)'+\mathcal T\, Y-\mathcal P y^+=0\,,\\
    &(x^+)'+\mathcal T_{(1)} Y-\mathcal Px^++\mathcal Ty_0=0\,.
\end{align}
\end{subequations}
Apart from these equations we have $Y'=0$. On-shell we have only three free functions $x_0$, $x_1$ and $y_1$ in the boundary conditions for the $\mathcal B$ field which couple to three functions $\mathcal P$, $\mathcal T$ and $\mathcal T_{(1)}$ in the gauge field invariantly in the boundary term. 
\subsection{Asymptotic symmetries}\label{subsec42}
Before shifting gears to our main objective which is finding the boundary action by making the action well-defined based on our adopted boundary condition, we pause here and ask what kind of symmetries we should expect to be realized by the boundary system. In other words, what is the symmetry algebra that governs the phase space of our theory? In order to answer this question, we study asymptotic symmetries of the bulk theory in its first-order formulation. We generate a general gauge transformation with the gauge parameter $\Lambda$ as in \eqref{generalgaugetrans},
\begin{equation}\label{BCPGT2}
    \Lambda=\laa\, J+\varepsilon^+ P_{+}+\varepsilon P_-+\laa_1J^{(1)}+\varepsilon_1^+P_+^{(1)}+\varepsilon_1P_-^{(1)}+\sigma J^{(2)}\,,
\end{equation}
and require that the boundary conditions on the 1-form connection $\mathcal A$ 
in its gauge fixed form, \eqref{L2bc} are preserved i.e. 
\begin{align}\label{BCPGT1}
\extd\Lambda+[\mathcal A,\Lambda]=\mathcal{O}(\delta \mathcal A)\,.
\end{align}
The components of $\Lambda$ are in one to one correspondence with those introduced for the scalar field $\mathcal B$ in \eqref{Bcompn}. {To a given gauge transformation $\Lambda$, using the bilinear form \eqref{genbilin}, we can associate an invariant pairing between the gauge transformation and all possible variations $\delta\mathcal A$ 
\begin{align}\label{charge2221}
\delta  {\mathcal C}[\Lambda]=-\kappa\int\langle  \Lambda,\delta\mathcal A\,\rangle_{\Gamma}\,.
\end{align}
In particular, if we implement the definition of \eqref{charge2221} to any transformation along $P_+$ we have $\mathcal C[\varepsilon^+]=0$ which shows that any transformation along $\varepsilon^+$ does not change the physical state. Since according to the gauge transformation condition \eqref{BCPGT1}, $\mathcal P_{(1)}$ shift transforms under $\epsilon^+$;
\begin{align}
    \delta_\Lambda \mathcal P_{(1)}=\lambda_1'+\mathcal T\varepsilon-\varepsilon^+\,,
\end{align}
 the gauge choice \eqref{gaugefixingP1} does not remove any physical configuration and is accessible provided that $\varepsilon^+=\lambda_1'+\mathcal T\varepsilon$. }
Preserving the boundary conditions \eqref{L2bc} demands that these parameters should satisfy the analogous equations as in \eqref{eomB1} and \eqref{eomB2} 
\begin{align}\label{parameters}
    \laa&= \varepsilon\mathcal P+\varepsilon'\,,\qquad\qquad\quad\;\,\laa_1=\mathcal P \varepsilon_1+\varepsilon_1'\,,\nn\\
\varepsilon^+&=\laa_1'+\mathcal T \varepsilon\,,\qquad\qquad\quad\varepsilon_1^+=\sigma'+\mathcal T_{(1)} \varepsilon+\mathcal T \varepsilon_1\,.
\end{align}
After plugging in the parameters \eqref{parameters} into \eqref{BCPGT2} we can read off the transformation induced by $\Lambda$ on the state-dependent functions from \eqref{BCPGT1}
\begin{align}\label{transWVir2}
    \delta_\Lambda\mathcal{P}&=\left(\varepsilon\mathcal P+\varepsilon'\right)'\,,\nn\\
    \delta_\Lambda\mathcal{T}&=\varepsilon\mathcal T'+2\varepsilon'\mathcal T+\varepsilon_1\left(\mathcal P'-\tfrac12\mathcal P^2\right)'+2\varepsilon_1'\left(\mathcal P'-\tfrac12\mathcal P^2\right)+\varepsilon_1'''\,,\nn\\
     \delta_\Lambda\mathcal{T}_{(1)}&=\big(\mathcal T_{(1)} \varepsilon+\mathcal T\varepsilon_1+\sigma'\big)'+\varepsilon'\mathcal T_{(1)}-\sigma'\mathcal P+\varepsilon_1'\mathcal T
     \,.
\end{align}
The transformation \eqref{transWVir2} manifests the expectation that $\mathcal P$, $\mathcal T$  and $\mathcal T_{(1)}$ under $\varepsilon$ are  primary fields of conformal weight (1, 2, 2) respectively. 

\subsection{Symmetry algebra and the coadjoint representation}\label{chargealgebra}

The transformation \eqref{transWVir2} are equivalent to the linearized form of the coadjoint representation of a centrally extended group of symmetries.\footnote{The non-linear form of the coadjoint representation can be identified in section \ref{sec5} when we derive the boundary action of the theory on the coadjoint orbit of the group. The coadjoint action is proportional to the Hamiltonian generator on the circle \cite{Stanford:2017thb,Alekseev:1988ce,Witten:1988xj}.} In fact, the phase space of the theory coincides with the coadjoint representation of the asymptotic symmetry group. In principle, these symmetries describe the Hilbert space of the theory around a vacuum solution which itself is labeled by constant representatives of the coadjoint orbit of the group.

To form the algebra of the coadjoint representation which coincides with the algebra of asymptotic symmetries, we use the invariant bilinear form of the theory as the pairing between the adjoint and coadjoint elements \eqref{charge2221}
\begin{align}\label{charge2}
\delta  {\mathcal C}[\Lambda]=\delta  {\mathcal C}[\varepsilon]+\delta {\mathcal C}[\varepsilon_1]+\delta {\mathcal C}[\sigma]\,,
\end{align}
where $\delta {\mathcal C}$'s on the right hand side of \eqref{charge2} can be determined by inserting the form of $\Lambda$ from \eqref{BCPGT2} and $\delta \mathcal A$ from \eqref{L2bc} into \eqref{charge2221} and implementing the bilinear form \eqref{bilinearL2}.
We have 
\begin{subequations}\label{generators123}
\begin{align}
{\mathcal C}[\varepsilon]&=\kappa\int\varepsilon\left(\mathcal T_{(1)}+\gamma_1\mathcal T-\tfrac{\gamma_0}{2}[\cP^2-2\cP']\right)\,,\\
{\mathcal C}[\varepsilon_1]&=\kappa\int\varepsilon_1\left(\mathcal T-\tfrac{\gamma_1}{2}[\cP^2-2\cP']\right)\,,\\
{\mathcal C}[\sigma]&=-\kappa\int\sigma\cP\,.
\end{align}
\end{subequations}
The pairing \eqref{charge2} does not coincide with the conventional notion of surface charges in BF-theories \cite{Grumiller:2015vaa,Grumiller:2017qao} that are defined on codimension-2 surfaces which in this case is a point --- see subsection \ref{subsec73}. However, by Wick rotating the boundary coordinate and considering the theory in Euclidean signature we can also think of the pairing as some unconventional finite and integrable  charges  over a space-like slice \cite{Afshar:2015wjm} {as they lead to an algebra (shown below) associated with the coadjoint representation of asymptotic symmetries of the Euclidean theory; in other words, the symmetry of the phase space of the theory. This correspondence between the phase space of the theory and the coadjoint representation of the group of symmetries acting on the boundary is specially justified in this case as the bulk piece of the action is zero and the BF-theory is equivalent to the boundary action as will be discussed later}. 

The compactified Euclidean circle then allows us to define the generators $L_m$, $T_m$ and $P_m$ as the Fourier modes of ${\mathcal C}[\varepsilon]$, ${\mathcal C}[\varepsilon_1]$ and ${\mathcal C}[\sigma]$ respectively. Using the relation $[{\mathcal C}_1,{\mathcal C}_2]=\delta_2 {\mathcal C}_1=-\delta_1{\mathcal C}_2$ we can find the algebra 
\begin{align}\label{WValgebra}
  &[L_n,L_m]=(n-m)L_{n+m}+\gamma_0n^3\delta_{n+m,0}\,,\nn\\\vspace{9mm}
  &[L_n,T_m]=(n-m)T_{n+m}+\gamma_1n^3\delta_{n+m,0}\,,\nn\\\vspace{9mm}
  &[L_n,P_m]=-mP_{n+m}+ i\kappa \,n^2\delta_{m+n,0}\,,\nn\\
 &[T_n,T_m]=(n-m)M_{n+m}+\frac{c}{12}n^3\delta_{n+m,0}\,,\nn\\
  &[T_n,P_m]=0\,, \qquad[P_{m},P_{n}]=0\,,
  \end{align}
  with
\begin{align}\label{twistedsug}
M_n=\sum_q P_{n-q}P_q+2i\kappa\, nP_{n}\,, \qquad\text{and}\qquad \kappa^2=\frac{c}{24}\,.
\end{align}
The algebra \eqref{WValgebra} is the spin-2 extension of the twisted warped conformal symmetry at vanishing Kac-Moody level \cite{Afshar:2015wjm,Afshar:2019axx,Afshar:2019tvp}. One can also read the algebra \eqref{WValgebra} as the spin-2 extension of the BMS$_3$ algebra using the prescription in \cite{Afshar:2015wjm}; one can consider the twisted Sugwara generator $M_n$ in \eqref{twistedsug} as independent and rewrite the algebra \eqref{WValgebra} in terms of $M_n$ where the only non-zero commutator that is altered is the third commutator replaced with
\begin{align}
    [L_n,M_m]=(n-m)M_{n+m}-\kappa\,n^3\delta_{m+n,0}\,.
\end{align}
{The appearance of a quadratic term in the commutator of two spin-2 generators $[T,T]$ in the CFT language is a feature of including  primaries of conformal dimensions $2, 3, \cdots$ corresponding to higher spin fields $s\geq2$ in the bulk. This implies that in general the algebra of higher spin gravity closes only in the so-called enveloping algebra such as $W$-algebras of higher spin theories in three dimensions \cite{Campoleoni:2010zq,Afshar:2013vka,Gonzalez:2013oaa}.}

\section{Variational principle}\label{varp}
The variation of the BF-action on-shell is a boundary term
\begin{align}\label{variationalprinc}
    \delta I\approx\kappa\int \extd u\,\langle\mathcal  B,\delta \mathcal A_u\rangle\,.
\end{align}
In order to have a well-posed variational principle we need to add a new boundary term to the action whose variation cancels the boundary term \eqref{variationalprinc}. 

\subsection{Boundary action at level-1}
First, we explain the general procedure for obtaining the boundary action at level-1 which is the BF-theory corresponding to the (twisted-)$\widehat{\text{CGHS}}$ model \eqref{BFseries1} with the Maxwell gauge algebra given in \eqref{Maxwel1} and the invariant bilinear form  \eqref{gggggg}-\eqref{JJJJJJ} with $\gamma_1=1$. The boundary condition for the gauge field and the  dilaton is \cite{Afshar:2019axx}:
\begin{align}\label{level1AB}
   \mathcal A_u=\cP\,J+\cT\,P_{+}+P_{-}\,,\qquad
\mathcal B=Y J+x_1P_-+x^+ P_++x_0J^{(1)}\,.
\end{align}
The boundary term \eqref{variationalprinc} at this level is
\begin{equation}\label{integrMAx}
    \delta I_{(1)}
    \approx \kappa\int \extd u\big[(\,x_0+\gamma_0\,Y\,)\delta\cP-\,x_1\,\delta\cT\big]\,,
\end{equation}
where $\gamma_0$ is a free parameter and all other variables under the integral are functions of $u$.
This boundary term which in general is not integrable should be canceled by variation of a supplemented boundary term to the BF-action. In order to make the boundary term \eqref{integrMAx} integrable we exploit the fact that our theory has two constants of motion at this level
\begin{equation}\label{casimirs}
    C_0\equiv Y\qquad\qquad  C_1\equiv\frac{1}{2}\langle \mathcal B,\mathcal B\rangle_{\gamma_1}=x_0Y-x^+x_1\,.
\end{equation}
Upon using the field equations $
    x^+=x_0'+\mathcal T x_1$ and $Y=x_1'+\mathcal P x_1
$ we can substitute $\mathcal T$ and $\mathcal P$ in terms of the two Casimirs \eqref{casimirs} into the boundary term \eqref{integrMAx}. After some algebra we get
\begin{align}\label{bndry1}
    \delta I_{(1)}&\approx
    \kappa\int \extd u\Big[\delta\frac{C_1}{x_1}+C_1\delta\frac{1}{x_1}-C_0\delta\frac{x_0}{x_1}+\delta x_0^\p-[x_0\delta\ln x_1]^\p\nn\\
    &\qquad \,\;\qquad+\tfrac12\gamma_0\Big(\delta\frac{C_0^2}{x_1}+C_0^2\delta\frac{1}{x_1}-2C_0[\delta \ln x_1]'\Big)\Big]\,.
\end{align}
The boundary term \eqref{bndry1} is integrable providing that the following holds
\begin{align}\label{integrab11}
   \delta\int\frac{\extd u}{x_1}=0\,,\qquad \delta\int\frac{x_0}{x_1}\extd u=0\,.
\end{align}
These integrability conditions will be interpreted in section \ref{sec5}, where we consider the theory on the circle, as having fixed zero mode for the Euclidean fields on the circle. It is worth mentioning that both of the conditions \eqref{integrab11} are required for a vanishing boundary term.\footnote{In \cite{Afshar:2019axx} only the first integrability condition was needed as $g\equiv\int^u x_0/x_1$ was considered to be a periodic function on the circle with no winding modes. Here we  relax this boundary condition but we have to fix the winding instead.
}

As a consequence of \eqref{integrab11} we can add the following boundary action to the BF-theory at this level to cancel the boundary term \eqref{bndry1};
\begin{align}\label{bndyaction1}
 I_{(1)}\to I_{(1)} + \partial I_{(1)}\,,\qquad  \partial I_{(1)}=-\kappa\int\frac{\extd u}{x_1}\left(C_1+\tfrac12\gamma_0 C_0^2\right) \,.
\end{align}
{Imposing the field equations, the bulk contribution vanishes and we are left with the boundary term $\partial I_{(1)}$.} In comparison with the boundary action for the $\widehat{\text{CGHS}}$ model  \cite{Afshar:2019axx} where only $C_1$ contribute, we see that here the Casimir $C_0$ can also contribute to the boundary action of the twisted-$\widehat{\text{CGHS}}$ model \eqref{BFseries1}, thanks to the invariant-bilinear form \eqref{gggggg}. In the Euclidean description of the theory in section \ref{sec5} this action leads to a  Warped-Schwarzian theory \cite{Afshar:2019tvp} with vanishing $u(1)$ level where $C_0^2$  contributes to the Schwarzian and $C_1$ to a twisted warped term.

\subsection{Boundary action at level-2}
Using the boundary condition for the gauge field and the dilaton given in \eqref{L2bc} and \eqref{Bcompn}, 
the boundary term of the action \eqref{variationalprinc} at this level {follows:}
\begin{equation}\label{bndry22}
    \delta I_{(2)}\approx\kappa\int \extd u \big[\gamma_0\,Y\,\delta \mathcal{P} +\gamma_{1}(y_0\,\delta \mathcal{P} -y_1\,\delta \mathcal{T} )+x_0\delta\cP-x_1\delta\cT-y_1\delta\cT_{(1)}\big]\,.
\end{equation}
There are three conserved quantities at this level, one linear and two bilinear ones associated to the algebra Casimirs \eqref{casimir222}
\begin{subequations}\label{casimir22}
\begin{align}
    C_0&\equiv Y\,,\\
    C_1&\equiv\frac{1}{2}\langle \mathcal B,\mathcal B\rangle_{\gamma_1}=y_0Y-y^+y_1\,,\\ 
    C_2&\equiv\frac{1}{2}\langle \mathcal B,\mathcal B\rangle_{\gamma_2}=x_0Y-y^+x_1-x^+y_1+\tfrac{1}{2}y_0^2\,.
\end{align}
\end{subequations}
We use  identities \eqref{casimir22} to eliminate the dilaton components $x^+$, $y^+$ and $Y$ in terms of Casimirs. Plugging them into field equations \eqref{eomB1} we trade $\cT$, $\mathcal P$ and $\mathcal T_{(1)}$ in the boundary term \eqref{bndry22} with independent dilaton fields $x_0$, $x_1$, $y_1$ and  Casimirs, then
substitute them into the boundary term \eqref{bndry22}. After doing some algebra we get\footnote{
To avoid messing up the calculation we plug in the dilaton component $y_0$ which is a  dependent field, only in the end;
\begin{equation}
      y_0=x_1'+\mathcal P x_1=\frac{x_1}{y_1}C_0+y_1\big(\frac{x_1}{y_1}\big)^\p\,.
\end{equation}
}
\begin{align}\label{deltaS2}
 \delta I_{(2)}&\approx 
\kappa\int\extd u\Big[\delta\frac{C_2}{y_1}+C_2\delta\frac{1}{y_1}-\delta\frac{C_1x_1}{y_1^2}-C_1\delta\frac{x_1}{y_1^2}-C_0\delta\frac{x_0}{y_1}+\tfrac{1}{2}\delta\frac{C_0^2x_1^2}{y_1^3}+ \tfrac{1}{2}C_0^2\delta\frac{x_1^2}{y_1^3}\nn\\&\qquad\qquad\;+\tfrac12\gamma_0\Big(\delta\frac{C_0^2}{y_1}+C_0^2\delta\frac{1}{y_1}\Big)+\gamma_1\Big(\delta\frac{C_1}{y_1}+C_1\delta\frac{1}{y_1} -\tfrac{1}{2}\delta  \frac{C_0^2x_1}{y_1^2} -  C_0^2 \delta \frac{x_1}{y_1^2} \Big)\Big]\,,
\end{align}
where the total derivative terms have been thrown away.
The constraint which comes from the well-posed variational principle is that the following integrals should be fixed 
\begin{align}\label{integrabilitylevel2}
\delta\int\frac{1}{y_1}=0\,,\qquad
\delta\int\frac{x_1}{y_1^2}=0\,,\qquad \delta\int\frac{x_1^2}{y_1^3}=0\qquad\text{and}\qquad \delta\int\frac{x_0}{y_1}=0\,.
\end{align}
As a consequence we complement the action by a boundary action $ I_{(2)}\to I_{(2)} + \partial I_{(2)}$ as,
\begin{align}\label{boundaryactionextended}
    \partial I_{(2)}=-\kappa\int \frac{\extd u}{y_1} \Big(C_2-[\frac{x_1}{y_1}]C_1+\tfrac12 \big[\frac{x_1}{y_1}\big]^2C_0^2+\gamma_1 \big( C_1-\tfrac{1}{2}[\frac{x_1}{y_1}]C_0^2\big)+\tfrac{\gamma_0}{2}\, C_0^2\Big)\,.
\end{align}
{Again, when the field equations hold, the bulk action vanishes and the Casimir functions in the boundary action are constant in $u$.} In the next section we will discuss the boundary actions derived in this section in level-1 \eqref{bndyaction1} and level-2 \eqref{boundaryactionextended} on the circle. The integrability conditions \eqref{integrabilitylevel2} will be interpreted as fixing the zero modes of certain functions on the circle.
\section{Euclidean theory}\label{sec5}

In this section we aim to study the theory at finite temperature $T=1/\beta$. In order to do so we perform a Wick rotation and work in Euclidean signature with periodic time $\tau\sim\tau+\beta$
\begin{align}\label{wickrot}
    u\to iu\equiv\tau\,.
\end{align}
The Wick rotation \eqref{wickrot} entails using the Euclidean frame metric $\delta_{ab}$ 
with $a,b=1,2$ which does not distinguish between the up and down indices. Considering $\epsilon_{21}=1$ the Euclidean algebra with $a,b=1,2$ has exactly the same form as the Lorentzian and we do not need to change anything at the level of the algebra. For example, at level-1, we have
\begin{align}\label{Euclalg}
    [P_a,J]=\epsilon_{ab}P_b\qquad[P_a,P_b]=\epsilon_{ab}J^{(1)}\,,\qquad a,b=1,2\,.
\end{align}
All generators in \eqref{Euclalg} are Euclidean. The map from \eqref{Euclalg} to its Lorentzian partner \eqref{maxwell2} is then
\begin{align}
    P_2\to-iP_0\,,\qquad P_1\to P_1\,,\qquad J\to iJ\,,\qquad J^{(1)}\to-i J^{(1)}\,.
\end{align}
The same holds for the next level-2. Namely the extension of the Euclidean algebra is exactly the same as in \eqref{level2alg1} with $a,b=1,2$ and $\epsilon_{21}=1$. The map from Euclidean extended generators to the Lorentzian ones is then
\begin{align}
    P_2^{(1)}\to-iP^{(1)}_0\,,\qquad P_1^{(1)}\to P_1^{(1)}\,,\qquad J^{(2)}\to i J^{(2)}\,.
\end{align}
{However, in our Lorentzian analysis we used the algebra form in the light-cone gauge  \eqref{Maxwel1}-\eqref{level2alglc}. 
The same algebra can be used for Euclidean signature, with the identifications
\begin{align}
     P_{\pm}=(P_1\pm iP_2)/\sqrt{2}\,,\qquad P_{\pm}^{(1)}=(P_1^{(1)}\pm iP_2^{(1)})/\sqrt{2}\,,
\end{align}
accompanied with i-rescalings of $J$'s as mentioned above. Since the light-cone algebra is the same as in the Lorentzian case the bilinear form  {of} the Euclidean algebra also remains the same as in \eqref{gggggg}-\eqref{bilinearL2}. This assures that,}
had we started the analysis of section \ref{varp} in  Euclidean signature we would get the same result for the boundary terms \eqref{bndyaction1} and \eqref{boundaryactionextended} with 
the analytic continuation of the integrand using $\partial_u\to i\partial_\tau$. Note that since the connection 1-form $\mathcal A$ is covariant we just need to replace $\int\to\oint$.
\subsection{Euclidean action at level-1}
Using the prescription of Wick rotation described above, the Euclidean continuation of the boundary action  at level-1 \eqref{bndyaction1}  can be evaluated
\begin{align}\label{Euclidact10}
I_{(1)}^{\text{\tiny E}}
=-\kappa  \oint\frac{\extd\tau }{x_1} \big[ C_1{+}\tfrac12\gamma_0  C_0^2\big]\,,
\end{align}
where the Casimirs at level-1 are obtained from \eqref{casimirs} using \eqref{wickrot} as
\begin{align}\label{EuclidCasimir2}
 C_0&=i( x_1'-i{\mathcal P} x_1) \,,\nn\\
 C_1&=i x_0( x_1'-i{\mathcal P}  x_1)- x_1(i x_0'+{\mathcal T}  x_1)\,,
\end{align}
in which prime denotes a differentiation w.r.t. $\tau$. We introduce the following  field redefinition,
\begin{align}\label{qusifunc1}
    f'=\frac{1}{x_1}\,,\qquad g'=ix_0 f'\,.
\end{align}
This choice is justified by the fact that the integrability conditions \eqref{integrab11} here on the circle are translated as having fixed zero modes for functions $f$ and $g$. This means that they are either periodic or quasi-periodic functions of $\tau$ {with fixed winding number}.
Inserting \eqref{EuclidCasimir2} into \eqref{Euclidact10} we find the following expression in terms of variables \eqref{qusifunc1}
\begin{align}\label{effectiveactio12}
I_{(1)}^{\text{\tiny E}}
=\kappa  \oint\frac{\extd\tau }{f'} \Big(\mathcal T^{\text{\tiny eff}}+i\mathcal P g'+g''{+}\tfrac12\gamma_0 [\frac{f''}{f'}]^2\Big)\,,
\end{align}
where $\mathcal T^{\text{\tiny eff}}=\mathcal T{-} \tfrac12\gamma_0 [\mathcal P^2-2i\mathcal P ']$ is the effective energy density. We can rewrite the effective action \eqref{effectiveactio12} in terms of the inverse function $h=-f^{-1}$ and get\footnote{We drop the total double derivative terms. {The conventional Schwarzian derivative term $\big(\frac{h''}{h'}\big)'-\frac12\big(\frac{h''}{h'}\big)^2$ is equivalent to $-\frac12\big(\frac{h''}{h'}\big)^2$ under the integral.}}
\begin{align}\label{effectiveactio1}
I_{(1)}^{\text{\tiny E}}=\kappa  \oint\extd f  \Big(h'^2\mathcal T^{\text{\tiny eff}}{+}\tfrac{\gamma_0}{2} \big[\frac{h''}{h'}\big]^2+g'\big[i\mathcal P h'-\frac{h''}{h'}\big]\Big)\,.
\end{align}
{The field redefinition $h=-f^{-1}$ entails
 \begin{align}
     f'=\frac{1}{h'}\circ f\,,\quad f''=-\frac{h''}{h'^3}\circ f\,,\quad g'=\frac{g'}{h'}\circ f\,,\quad g''=\big(\frac{g''}{h'^2}-\frac{g'h''}{h'^3}\big)\circ f\,,\quad\cdots\nn
 \end{align}
It is understood that  in \eqref{effectiveactio1} ${}^\prime$ is with respect to the  variable $f$ and we drop $\circ f$ from here on.}
 
{The Euclidean action obtained in \eqref{effectiveactio1} is associated with} a warped-Schwarzian theory \cite{Afshar:2019tvp}  with vanishing $u(1)$ level which is an action for the warped Virasoro group coadjoint orbits \cite{Afshar:2015wjm}. {Here upon imposing suitable boundary conditions, we found this boundary action equivalent to the Euclidean twisted-$\widehat{\text {CGHS}}$ model \eqref{CGHShat}, whose  bulk contribution vanishes due to constraints. The same way as the Schwarzian action governs the theory of pseudo-Goldstone modes associated to a symmetry breaking of the Virasoro group to its $SL(2,\mathbb{R}$) or $U(1)$ subgroups, the action \eqref{effectiveactio1} determines the theory of Goldstone modes appearing in the symmetry breaking from the twisted warped Virasoro group at vanishing $U(1)$-level to its finite dimensional global subgroups $ISO(1,1)_c$ which is the 2D Maxwell group (the central extension of the 2D Poincar\'e group) or $U(1)\times U(1)$ depending on the value of zero-modes of $\mathcal{T}$ and $\mathcal{P}$.}

Field equations of the action  \eqref{effectiveactio1} after integrating once yields ---  from now on we restrict ourselves  to  constant representatives $\mathcal T_0$ and $\mathcal P_0$,
\begin{align}\label{Wschth}
   {-} i\mathcal P_0 h'+\frac{h''}{h'}&=a_0\,,\\
 2h'\mathcal T_0^{\text{\tiny eff}}{+}i\mathcal P_0g'+\frac{g''}{h'}{-}\frac{\gamma_0}{h'}\big(\frac{h''}{h'}\big)'&=b_0\,.\label{Wschth2}
\end{align}
where  $a_0$ and $b_0$ are integration constants. The general solution as well as the disk partition function for this model are obtained in \cite{Afshar:2019tvp}.\footnote{The cylinder partition function of the $\widehat{\text{CGHS}}$ model is considered in \cite{Godet:2020xpk}.} However for evaluating the on-shell action we do not need to solve these equations. The fact that in \eqref{effectiveactio1} we use $f$ as reparametrizing the angle coordinate on the circle implies that $h$ is a quasi-periodic function in $\beta$,
 \begin{align}\label{quasiperiodic}
     \oint h'=\beta\,.
 \end{align}
  Using \eqref{quasiperiodic} we can determine the value of $a_0$ by integrating once the equation \eqref{Wschth},
\begin{align}\label{integrationconst1}
    a_0={-}i\mathcal P_0\,.
\end{align}
Further, multiplying both sides of equation \eqref{Wschth} by $h'$ and integrating once  we get,
\begin{align}\label{integrationconst2}
   \oint h'^2=\beta \,.
\end{align}
 Using \eqref{Wschth} together with \eqref{integrationconst1} and \eqref{integrationconst2} we can evaluate the on-shell action
\begin{align}\label{onshell1}
 I^{\text{\tiny E}}_{(1)}|_{\text{\tiny on-shell}}=\kappa\beta(\mathcal T_0{-} \tfrac12\gamma_0 \mathcal P_0^2){+}i\kappa\mathcal P_0\oint g'\,.
\end{align}
The last term measures the quasi-periodicity of the function $g$ which according to the integrability condition \eqref{integrab11} has to be fixed and can be determined in terms of $b_0$, $\mathcal{T}_0$ and $\mathcal P_0$; substituting the $h''$ from equation \eqref{Wschth} into \eqref{Wschth2} we get a second order equation for $g$
\begin{align}\label{sndeqWsh}
    2h'\mathcal T_0{+}i\mathcal P_0g' +\frac{g''}{h'}{-}\gamma_{0}\mathcal{P}_{0}^2=b_0\,,
\end{align}
where upon integrating it once we have
\begin{align}\label{sndeqWsh3}
    0=\beta(2\mathcal T_0{-} \gamma_0 \mathcal P_0^2-b_0){+}i\mathcal P_0\oint g'+\oint\frac{g''}{h'}\,.
\end{align}
The last term in \eqref{sndeqWsh3} is zero because if we rewrite it in terms of the inverse map $h\to h^{-1}$ it is the integral on derivative of a periodic function
\begin{align}\label{gprimprimehprime}
     \oint\frac{g''}{h'}\to\oint \left(\frac{g'}{h'}\right)'=0\,.
\end{align}
Thus the value of the on-shell action \eqref{onshell1} and the corresponding free energy is  \begin{align}\label{onshell12}
 F \supset \beta^{-1} I^{\text{\tiny E}}_{(1)}|_{\text{\tiny on-shell}}=-\kappa(\mathcal T_0{-} \tfrac12\gamma_0 \mathcal P_0^2-b_0)\,.
\end{align}

\subsection{Euclidean action at level-2}
The Euclidean continuation of the boundary action \eqref{boundaryactionextended} is
\begin{align}\label{boundaryactionextended12}
  I^{\text{\tiny E}}_{(2)}={-}\kappa\oint \frac{\extd \tau}{y_1} \Big(C_2-[\frac{x_1}{y_1}]C_1+\tfrac12 \big[\frac{x_1}{y_1}\big]^2C_0^2{+}\gamma_1 \big( C_1-\tfrac{1}{2}[\frac{x_1}{y_1}]C_0^2\big)+\tfrac{\gamma_0}{2}\, C_0^2\Big)\,.
\end{align}
Where the Casimirs are Wick rotated
\begin{align}\label{CasimirEucl2}
C_0&=i( y_1'-i\mathcal P y_1)\,,\\
 C_1&=-(x_1'-i\mathcal P x_1)( y_1'-i\mathcal P y_1)+ y_1[( x_1'-i\mathcal P x_1)'-\mathcal T y_1]\,,\nn\\
 C_2&=i x_0( y_1'-i\mathcal P y_1)+ x_1[( x_1'-i\mathcal P x_1)'-\mathcal T y_1]- y_1(i x_0'+\mathcal T_{(1)} y_1+\mathcal T x_1)-\tfrac12(x_1'-i\mathcal P x_1)^2 \,.\nn
\end{align}
It is convenient to introduce the following field redefinition,
\begin{align}\label{newfields2}
    f'=\frac{1}{y_1}\,,\qquad w'=\frac{x_1}{y_1}=x_1f'\,,\qquad g'=ix_0 f'\,.
\end{align}
Upon inserting \eqref{newfields2} into the Casimir combinations appearing in \eqref{boundaryactionextended12} we have
\begin{align}
C_2-\big[\frac{ x_1}{ y_1}\big] C_1+\tfrac12 \big[\frac{ x_1}{ y_1}\big]^2C_0^2&=-\frac{1}{f'^2}\Big(i{\mathcal P}g'+g''+{\mathcal T}_{(1)}+{\mathcal T}w'+\tfrac12 w''^2\Big)\,,\nn\\
C_1-\tfrac{1}{2}\big[\frac{ x_1}{ y_1}\big]C_0^2&=-\frac{1}{f'^2}\Big({\mathcal T}-w'''+w''\frac{f''}{f'}+w'\big(\frac{f'''}{f'}-\tfrac32\frac{f''^2}{f'^2}+i{\mathcal P}'-\tfrac12{\mathcal P}^2\big)\Big)\,,\nn\\
 C_0^2&=-\frac{1}{f'^2}\Big(\frac{f''^2}{f'^2}+2i{\mathcal P}\frac{f''}{f'}-{\mathcal P}^2\Big)\,.
\end{align}
The Euclidean effective action at level-2 \eqref{boundaryactionextended12} after inserting above and performing  some partial integration is obtained
\begin{align}\label{effectiveEuclid2}
      I^{\text{\tiny E}}_{(2)}&=\kappa\oint \frac{\extd \tau}{f'}\Big(i{\mathcal P}g'+g''+{\mathcal T}_{(1)}^{\text{\tiny eff}}+{\mathcal T}^{\text{\tiny eff}}w'+\tfrac12 w''^2{+}\gamma_1w'\big(\frac{f'''}{f'}-\tfrac32\frac{f''^2}{f'^2}\big)+\frac{\gamma_0}{2}\frac{f''^2}{f'^2}\Big)
\end{align}
where\footnote{The format appearing as ${\mathcal T}^{\text{\tiny eff}}$ and $  {\mathcal T}_{(1)}^{\text{\tiny eff}}$ are the same as those appearing in the expression of charge \eqref{generators123} after we do the analytic continuation \eqref{wickrot}.}
\begin{align}\label{effectiveval}
    {\mathcal T}_{(1)}^{\text{\tiny eff}}&={\mathcal T}_{(1)}{+}\gamma_1{\mathcal T}+\gamma_0\big(i{\mathcal P}'-\tfrac12{\mathcal P}^2\big)\,,\\
    {\mathcal T}^{\text{\tiny eff}}&={\mathcal T}{+}\gamma_1\big(i{\mathcal P}'-\tfrac12{\mathcal P}^2\big)\,.
\end{align}
If we rewrite \eqref{effectiveEuclid2} in terms of the inverse map $h=-f^{-1}$ we get
\begin{align}\label{Euclidbndry2}
      I^{\text{\tiny E}}_{(2)}&=\kappa\oint \extd f\Big(h'^2{\mathcal T}_{(1)}^{\text{\tiny eff}}{+}\frac{\gamma_0}{2}\frac{h''^2}{h'^2}+g'\big[i{\mathcal P}h'-\frac{h''}{h'}\big]\nn\\
      &\qquad\qquad\qquad+\frac{w'}{h'}\big[{\mathcal T}^{\text{\tiny eff}}h'^2{-}\gamma_1\big(\frac{h'''}{h'}-\tfrac32\frac{h''^2}{h'^2}\big)\big]+\tfrac12 \big[\big(\frac{w'}{h'}\big)'\big]^2\Big)\,.
\end{align}
Prime here refers to derivative w.r.t. $f$ and again we restrict our selves to constant representatives $\mathcal T_0$, ${\mathcal T}_{(1)}{}_0$ and $\mathcal P_0$ on the orbit. Upon varying the action w.r.t. $g$, $w$ and  $h$ we get
\begin{align}\label{second-level-Eu1}
  &i{\mathcal P}_0h'-\frac{h''}{h'}=a_0\,,\\
  &{\mathcal T}^{\text{\tiny eff}}_0h'^2{-}\gamma_1\big(\frac{h'''}{h'}-\tfrac32\frac{h''^2}{h'^2}\big)+\big(\frac{w'}{h'}\big)''=b_0h' \label{second-level-Eu2}\,,\\
  &2h'{\mathcal T}_{(1)}^{\text{\tiny eff}}{}_0+w'{\mathcal T}^{\text{\tiny eff}}_0+i{\mathcal P}_0g'+\frac{g''}{h'}+\frac{w'}{h'^2}\big(\frac{w'}{h'}\big)''\nn\\
  &{+}\gamma_1\big[\frac{w'}{h'^2}\big(2\frac{h'''}{h'}+\tfrac32\frac{h''^2}{h'^2}\big)-(\frac{w'}{h'^2})''-\frac{3}{h'}(\frac{h''w'}{h'^2})'\big]{-}\frac{\gamma_0}{h'}(\frac{h''}{h'})'=c_0\,.\label{second-level-Eu3}
\end{align}
where $a_0$, $b_0$ and $c_0$ are integration constants. 

Following the same logic as in the level-1, by integrating equations \eqref{second-level-Eu1}-\eqref{second-level-Eu2} and using the fact that $h'$ is quasi-periodic in $\beta$  one obtains
\begin{align}\label{a0-P0}
   a_{0}= i{\mathcal{P}}_{0}\,,\qquad b_0={\mathcal T}^{\text{\tiny eff}}_0\,,
\end{align}
where in the last equation we used the identity \eqref{integrationconst2}. In fact using eq. \eqref{second-level-Eu1} together with the identity \eqref{integrationconst2} iteratively one can generalize the identity 
\begin{align}\label{integrationconst3}
    \qquad \oint (h')^n=\beta\,,\qquad n\geq1\,.
\end{align}
We can also derive same identity for $w'$ or $g'$ smeared with powers of $h'$. Using the inverse map $h\to h^{-1}$ we have $h'\to1/h'$ and $w'\to w'/h'$ which translate to
\begin{align}\label{identitynum1}
    \oint\frac{1}{(h')^n}=\oint (h')^n\,,\qquad\oint \frac{w'}{(h')^n}=\oint w'(h')^{n-1}\,,\qquad n\geq1
\end{align}
Under the inverse map $h\to h^{-1}$ we also have $w''\to \frac{1}{h'^2}\big(w''-\frac{h''}{h'}w'\big)$ which, upon using the fact that $\oint w''=0$ implies
\begin{align}\label{wprimehprime}
    \oint\frac{w''}{h'^2}=\oint w'\frac{h''}{h'^3}
    =i\mathcal P_0[\oint \frac{w'}{h'}-\oint \frac{w'}{h'^2}]=i\mathcal P_0[\oint w'-\oint w'h']\,,
\end{align}
where in the {second} equality we used \eqref{second-level-Eu1} and in the last equality we used the identity of \eqref{identitynum1}. On the other hand if we use the partial integration we have $\oint\frac{w''}{h'^2}=2\oint w'\frac{h''}{h'^3}$. This implies that the right hand side of \eqref{wprimehprime} is zero;
\begin{align}\label{ident876}
    \oint w'h'=\oint w'\,.
\end{align}
Using the same logic iteratively we get a generalization of the identity in \eqref{gprimprimehprime} and of \eqref{ident876}
\begin{align}\label{identitynum2}
    \oint\frac{w''}{(h')^n}={\oint w''(h')^{n}}=0 \,,\qquad \oint w'(h')^n=\oint w'(h')^{n-1}\,,\qquad n\geq0\,.
\end{align}
In order to find identities for higher powers of $w'$ we use \eqref{second-level-Eu2}. If we insert $a_0$ and $b_0$ from \eqref{a0-P0} and $h''$ from eq. \eqref{second-level-Eu1} we can simplify and reduce the order of derivatives in \eqref{second-level-Eu2} and find a third order differential equation for $w$
\begin{align}\label{second-level-Eu22}
  &{\mathcal T}_0h'(h'-1){+}\frac{\gamma_1}{2}{\mathcal P}_0^2(h'-1)+\frac{w'''}{h'}-i{\mathcal P}_0\big(h'-1\big)\big(2\frac{w''}{h'}+i{\mathcal P}_0\frac{w'}{h'}\big)=0\,.
\end{align}  
Next we use eqs. \eqref{second-level-Eu1} and \eqref{second-level-Eu22} to solve for $h''$ and $w'''$ and substitute them in eq. \eqref{second-level-Eu3}
\begin{align}
     &-2h'{\mathcal T}_{(1)}^{\text{\tiny eff}}{}_0{-}\gamma_0{\mathcal{P}}_{0}^2(h'-1)-i{\mathcal P}_0g'-\frac{g''}{h'}\nn\\
 & {-}\frac{\gamma_1}{h'}\big[{\mathcal T}^{\text{\tiny eff}}_0h'(h'-1){+}\frac{\gamma_1{\mathcal P}_0^2}{2}\big(h'^2-1\big) -i{\mathcal{P}}_{0}(h'-1)\frac{w''}{h'}-w'{\mathcal{P}}_{0}^2(h'-1)\big]-{\mathcal T}^{\text{\tiny eff}}_0\frac{w'}{h'}=c_0\label{second-level-Eu31}.
\end{align}
Integrating eq. \eqref{second-level-Eu31} once and using the identities in \eqref{identitynum1} and \eqref{identitynum2} we get,
\begin{align}\label{c0value1}
     -2\beta{\mathcal T}_{(1)}^{\text{\tiny eff}}{}_0-i{\mathcal P}_0\oint g' -{\mathcal T}^{\text{\tiny eff}}_0\oint w'=c_0\beta\,.
\end{align}
Finally one can obtain the on-shell action as 
\begin{align}\label{onshell123}
     I^{\text{\tiny E}}_{(2)}|_{\text{\tiny on-shell}}&=\kappa \Big( \beta{\mathcal T}_{(1)}^{\text{\tiny eff}}{}_0  +i{\mathcal P}_{0}\oint g'+{\mathcal T}^{\text{\tiny eff}}_0\oint  w' \Big)\\
     &={-}\kappa\beta\big({\mathcal T}_{(1)}{}_0{+}\gamma_1{\mathcal T}_0-\tfrac12\gamma_0{\mathcal P}_0^2{+}c_0\big)\,.\nn
\end{align}
where in the first line we used the identities \eqref{identitynum1} and \eqref{identitynum2} and in the second line we substituted  from \eqref{c0value1} and zero modes in \eqref{effectiveval}. 

\section{Thermodynamics}\label{sec7}
In this section we analyze the disk thermodynamics associated to the zero-mode solution represented by $\mathcal T_0$, $\mathcal P_0$ and $\mathcal T_{(1)}{}_0$ in the Euclidean gauge field,
\begin{align}\label{gaugefieldeuclid}
  i \mathcal A_\tau=\mathcal P_0\,J+\mathcal T_0\,P_{+}+P_-+\mathcal T_{(1)}{}_0\,P_+^{(1)}\,.
\end{align}
where all generators belong to the Euclidean algebra. The gauge field \eqref{gaugefieldeuclid} defines our boundary conditions in the extended-$\widehat{\text{CGHS}}$ model based on the BF-theory formulation of the extended Poincar\'e algebra at level-2. At level-1, the  gauge field has no components along $P_+^{(1)}$, hence $\mathcal T_{(1)}{}_0=0$. The level-1 BF-theory  formulation which is the description of the twisted-$\widehat{\text{CGHS}}$ model, is based on the  centrally extended  Poincar\'e algebra.
\subsection{Holonomy}
To ensure the regularity of the zero-mode solution on the disk, we require that the holonomy is trivial. This condition suggests that it belongs to the center of the group. The holonomy of $\mathcal A_\tau$ along the thermal cycle is
\begin{align}\label{holonomy1}
    \text{Hol}(\mathcal A_\tau)=\exp{\big[ \oint \mathcal A_\tau}\big]\in Z(G)\,.
\end{align}
 This ensures having contractible thermal cycles on the disk \cite{Bunster:2014mua}. The detailed analysis of finding the center of the group is made by applying the lemma to the Baker–Campbell–Hausdorff formula in the appendix \ref{centerapp}. Here we just report the result. The holonomy \eqref{holonomy1} belongs to the center of the group providing that at level-1 and level-2 extensions of the gauge field we have
 \begin{align}\label{holonomy}
\begin{array}{lll}
    \text{Level-1}:\qquad\qquad &  \mathcal P_0={\dfrac{2\pi n}{\beta}}\,,
\\[2.5ex]
     \text{Level-2}:\qquad\qquad &  \mathcal P_0=\dfrac{2\pi n}{\beta}\,, \qquad\mathcal T_0=0\,.
  \end{array}
   \end{align}
where there is no constraint on $\mathcal T_0$ at level-1 and no constraint on $\mathcal T_{(1)}{}_0$ at level-2. 

\subsection{Entropy from the onshell action}
The free energy of the system is $F=-T\ln Z$ where $Z$ is the partition function of the theory. At the classical level we have $Z=e^{-I_{\text{\tiny E}}}$, where $I_{\text{\tiny E}}$ refers to the Euclidean on-shell action. The corresponding entropy is then
\begin{align}\label{entropyonshell}
    S=-\frac{\partial F}{\partial T} =-I_{\text{\tiny E}}-T\frac{\partial I_{\text{\tiny E}}}{\partial T}\,.
\end{align}
Using the expression for the on-shell action \eqref{onshell1} for the twisted-$\widehat{\text{CGHS}}$ model and imposing the holonomy condition $\mathcal P_0=2\pi/\beta$ the corresponding entropy is obtained
\begin{align}\label{StwCGHS}
   S_{\text{\tiny tw-}\widehat{\text{\tiny CGHS}}}=2\pi\kappa\Big(-\frac{\mathcal T_0}{\mathcal P_0}+ x_0{+}\gamma_0 {\mathcal P_0}\Big) =2\pi\kappa \big(X_{\text{\tiny H}}{+}\gamma_0Y\big)\,.
\end{align}
The last equality is written in terms of the value of the dilaton field \eqref{Xdilaton} at the horizon  {$X_{\text{\tiny H}}= x_0+r_{\text{\tiny H}} x_1$ where $r_{\text{\tiny H}}=-\frac{\mathcal T_0}{\mathcal P_0}$ while $ x_0$ and $ x_1$ here are constant and counts the winding numbers associated to $g$ and $h$ fields respectively. According to \eqref{qusifunc1} we have;
\begin{align}
     x_0=-\frac{i}{\beta}\oint \frac{g'}{f'}=-\frac{i}{\beta}\oint g'\,,\quad\qquad x_1=\frac{1}{\beta}\oint\frac{1}{f'}=\frac{1}{\beta}\oint h'=1\,,
\end{align}
where in the last equality we used the field redefinition $h=-f^{-1}$ and used the identities \eqref{integrationconst1} and \eqref{ident876}.}
In \eqref{StwCGHS} we have also used the fact that $ x_1\mathcal P_0=Y$.
The entropy of the black hole in the twisted-$\widehat{\text{CGHS}}$ model in comparison to the $\widehat{\text{CGHS}}$ model acquires a temperature-dependent term. Its specific heat  in contrast to the $\widehat{\text{CGHS}}$ model, is then finite and linear in $T$;
\begin{align}
    C=T\frac{\partial S}{\partial T}=(2\pi)^2\kappa\gamma_0 T\,.
\end{align}
At level-2 we use the expression of the on-shell action  for the extended-$\widehat{\text{CGHS}}$ model in \eqref{onshell123} and also impose the holonomy condition \eqref{holonomy} which on the disk entails $\mathcal P_0=2\pi/\beta$ and $\mathcal T_0=0$. The entropy is then found from \eqref{entropyonshell}
\begin{align}\label{exCGHSentropy}
S_{\text{\tiny ex-}\widehat{\text{\tiny CGHS}}}&=2\pi\kappa\Big(-\frac{\mathcal T_{(1)}{}_0}{\mathcal P_0}+{ x}_0{+}\gamma_1{ x}_1 \mathcal P_0 +\gamma_0\mathcal P_0\Big)\nn\\
 &=2\pi\kappa\Big(-\frac{\mathcal T_{(1)}{}_0}{\mathcal P_0}+
  X_{\text{\tiny H}}{+}\gamma_1 Y_{(1)}{}_{\text{\tiny H}}+\gamma_0Y\Big)\,.
\end{align}
{In the first line upon using \eqref{newfields2} we have $x_0=-\frac{i}{\beta}\oint g'$ and $ x_1=\frac{1}{\beta}\oint w'$ counting the winding numbers of $g$ and $w$ quasi-periodic functions correspondingly and} the last equality is written in terms of the horizon value of dilaton fields \eqref{Ydilaton} and \eqref{Xdilaton} present in the extended-$\widehat{\text{CGHS}}$ model {(note that in this case $r_{\text{\tiny H}}=0$ as $\mathcal T_0=0$ on the disk)}. The first term in the expression of the entropy {can be} identified with the contribution of the spin-2 gauge field $f_{uu}$ in \eqref{fsolution} at the horizon.
{
\subsection{Entropy from canonical charges}\label{subsec73}
In BF theories \eqref{BFtheoryaction}  the gauge transformations $\mathcal A\to \mathcal A+\extd\Lambda$ could lead to codimension-2 canonical surface charges which in this case is a point \cite{Grumiller:2015vaa,Grumiller:2017qao}
\begin{align}
\delta Q[\Lambda]=\kappa\langle \Lambda, \delta \mathcal B \rangle
\end{align}
In particular, let us evaluate this expression at the horizon in \eqref{EF metric} for the null Killing vector $\xi=\partial_u$ to find the entropy as a Noether conserved quantity using the first law $\delta Q = T \delta S$. The associated gauge transformation is related to infinitesimal diffeomorphisms as $\Lambda=\xi^{\mu}\mathcal A_{\mu}=\mathcal{A}_u$. Thus we have
\begin{align}\label{TdS}
T\delta S=\kappa\langle \mathcal A_u, \delta \mathcal{B} \rangle\,.
\end{align}
At level-1, the zero-mode solution is given by \eqref{level1AB} where upon substituting $x_1=1$ and $\mathcal P_0=Y=2\pi T$ as fixed quantities (solving the dilaton field equation $x_1'+\mathcal Px_1=Y$) we have,
\begin{align}
T\delta S=2\pi\kappa T\left(\gamma_0\delta Y+\delta x_0-\delta (\mathcal T_0/\mathcal P_0)\right) \,,
\end{align}
which is an integrable identity and after integration leads to $S=2\pi\kappa(X_{\text{\tiny H}}+\gamma_0Y)$ in agreement with \eqref{StwCGHS}. At level-2 by inserting  zero mode components \eqref{L2bc} and \eqref{Bcompn} into equation \eqref{TdS} and using field equations \eqref{eomB1} to fix $y_1=1$ and $y_0=x_1{\mathcal P}_0$, we get 
\begin{align}\label{zeromodeBcompn2}
T\delta S=2\pi\kappa T\left(-\delta({\mathcal T}_{(1)}{}_0/{\mathcal P}_0)+\delta (x_0+x_1 r_{\text{\tiny H}})+\gamma_1\delta (y_0+y_1 r_{\text{\tiny H}})+\gamma_0 \delta Y\right) \,,
\end{align}
which upon integrating we find the same expression for the entropy as in \eqref{exCGHSentropy}.
}

\section{Summary and remarks}\label{sec8}
In this paper, we studied holographically the two-dimensional dilaton
gravity coupled to spin-2 gauge fields in flat spacetime. In particular we used the expansion method around the flat background $R=0$ to study interacting massless spin-2 theories and their holographic picture. We called this model as extended-$\widehat{\text{CGHS}}$ model. The action corresponds to a BF-theory for a 7-dimensional gauge group. Four of its generators comprise the centrally extended 2D Poincar\'e subalgebra $J,P_a,J^{(1)}$ and the two extra generators $P_a^{(1)}$ are associated with the spin-2 extension manifested as the spin-2 representation of the boost generator, and there exists a central generator $J^{(2)}$. The presence of the central term is assuring the existence of a non-degenerate bilinear form. We also derived the metric formulation of the theory and the corresponding field equations and solved them with appropriate gauge fixing/boundary conditions in section \ref{sec3}.

By translating the boundary conditions to the BF-formulation, we studied the asymptotic symmetries of the model in section \ref{sec4} which is the spin-2 extension of the warped-Virasoro algebra at vanishing Kac-moody level. In section \ref{varp} using the BF-theory setup and our adopted boundary conditions we studied the variational principle and found the appropriate boundary term which makes the variational-principle well-defined. This amounted to impose certain integrability conditions on the dilaton fields in the theory. We analytically continued the boundary action --- which equals the bulk action --- to the Euclidean signature where the integrability conditions gain simple forms. These integrability conditions amount to fixing the zero-mode of certain quasi-periodic functions which themselves account for a certain combination of dilaton fields components on the circle. The Euclidean boundary action on the circle together with the field equations are derived in terms of these quasi-periodic functions in section \ref{sec5} which is the spin-2 extension of the warped-Schwarzian theory at vanishing $U(1)$-level \cite{Afshar:2019tvp,Afshar:2019axx}. Finding the on-shell action and imposing the holonomy conditions on the disk we explored the thermodynamics of black holes in this setup and found the corresponding entropy in terms of the contribution of the spin-2 and dilaton fields at the horizon in section \ref{sec7}.

Based on our results in this paper both on the gravity side and on the Euclidean quantum mechanics side one can ask several questions. One of the interesting questions is whether there exists an SYK-like statistical quantum mechanics whose low-energy effective action is \eqref{Euclidbndry2}.  In fact, it was shown in \cite{Afshar:2019axx} that there exists a scaling limit from the effective action of the charged SYK model to the warped Schwarzian action \eqref{effectiveactio1} in which, in addition to the reparametrization fields $h$ comprising the Schwarzian term, the spin-1 fields $g$ are also present. In the spin-2 extended version \eqref{Euclidbndry2} the analogous scaling limit if exists would require starting from a charged SYK model which exhibits an extra spin-2 global symmetry. Another interesting calculation is the partition function of this model \eqref{Euclidbndry2} which will be 1-loop exact as its Schwarzian \cite{Stanford:2017thb} and warped Schwarzian \cite{Afshar:2019tvp} ancestors. 

From the bulk point of view, the appearance of extra spin-2 gauge symmetry can be considered as a blessing to flat-space holography in 2D. In the presence of extra massless gauge fields the topology of the group structure of the extended Poincar\'e symmetry is more involved and the number of configurations with $R=0$ enhances. In the pure dilaton-gravity in fact the constraint $R=0$ allows only very few saddles to the theory. Now with the appearance of extra spin-2 gauge symmetry, this can in principle enhance. This may increases the chance of an ensemble interpretation of the theory \cite{Saad:2019lba}. Similar to three dimensional higher-spin gravity \cite{Campoleoni:2010zq}, interesting features may also arise due to extension of the symmetry algebra; the notion of inequivalent non-trivial embeddings of the gravitational subalgebra in the full gauge algebra \cite{deBoer:1993iz,Campoleoni:2011hg,Gutperle:2011kf,Gary:2012ms,Afshar:2012nk,Afshar:2012hc} or the appearance of generalized black holes see e.g. \cite{Castro:2011fm,Perez:2012cf,deBoer:2013gz,Bunster:2014mua}.
\subsection*{Acknowledgement}
{We thank the anonymous referee for his/her careful reading  of our manuscript and questions which led to clarify some parts.} HA thanks Blagoje Oblak for fruitful discussions. He also thanks Hern\'an Gonz\'alez, Daniel Grumiller, and Dmitri Vassilevich for their collaboration on flat space holography in 2D dilaton-gravity. {ES and HRS acknowledge the support of Saramadan grants ISEF/M/99376 and ISEF/M/99407.}
\appendix

\section{Level-\texorpdfstring{$\infty$}{infty} theory}\label{Appendix1}

In section \ref{sec2}, we discussed the extension of the BF-theory to an arbitrary level-$N$. In the limit $N\to\infty$,  infinitely many copies of spin-2 fields can be summed into smooth functions of the expansion parameter $\sigma$ and comprise an effective 2-dimensional geometry with an effective curvature $\sigma$ as will be discussed here.

 \subsection{Defining summed variables}
 In order to build a BF-theory based on the infinite-dimensional Lie algebra \eqref{infinite-poinccare}, we extend the definitions \eqref{bfgaug2} to include infinitely many terms while introducing a regulating parameter $\sigma$:
\begin{subequations}
\begin{align}
    \mathbf{e}&= \sum_{k=0}^\infty \frac{\sigma^k}{k!} e_{(k)}^a(x){P}_a^{(k)}\,,\\
    \boldsymbol{\omega}&=\sum_{k=0}^\infty \frac{\sigma^k}{k!} \omega_{(k)}(x){J}^{(k)}\,.
\end{align}
\end{subequations}
Suppose that there are smooth one-forms $e^a(x,\sigma)$ and $\omega(x,\sigma)$ with convergent Taylor series  in $\sigma$ whose coefficients are given by $e^a_{(k)}$ and $\omega_{(k)}$. We have $ \mathbf{e}= \mathbf{e}[e^a(x,\sigma)]$ and $ \boldsymbol{\omega}= \boldsymbol{\omega}[\omega(x,\sigma)]$. The commutator of Lie-algebra-valued one-forms are
\begin{align}\label{eecom}
    \big[\mathbf{e}[e^a(x,\sigma)],\mathbf{e}[f^a(x,\sigma)]\big]&=\sum^\infty_{k,m=0}J_{k+m+1}\frac{\sigma^{k+m}}{k!m!}\epsilon_{ab}e^a_{(k)}(x)f^b_{(m)}(x)=\sum^\infty_{s=0}J_{(s+1)}\frac{\sigma^s}{s!}(e^af^b\epsilon_{ab})_{(s)}(x).
\end{align}
In the last equality, we used $ (ef)(x)=\sum^\infty_{k=0}\frac{1}{k!}\sum_{m=0}^ke_{(m)}(x)f_{(k-m)}(x)\binom{k}{m}$. 
The right-hand-side of \eqref{eecom} can be resummed as
\begin{align}
     \big[\mathbf{e}[e^a(x,\sigma)],\mathbf{e}[f^a(x,\sigma)]\big]&= \boldsymbol{\omega}[\epsilon_{ab}\int_0^\sigma(e^af^b)(x,\sigma^\p)d\sigma^\p]\,.
\end{align}
Similarly, we have
\begin{equation}
      \big[\mathbf{e}[e^a(x,\sigma)],\boldsymbol{\omega}[\omega(x,\sigma)]\big]= \mathbf{e}[\epsilon_{ab}(\omega \,e^b)(x,\sigma)]\,.  
\end{equation}
Now we are ready to define the Level-$\infty$ BF-theory by introducing the gauge field
\begin{equation}\label{infgauge}
    \mathcal{A}=\mathbf{e}[e^a(x,\sigma)]+ \boldsymbol{\omega}[\omega (x,\sigma)]\,,
\end{equation}
with curvature
\begin{align}
  \mathcal{F}&=\mathbf{e}[\extd e^a(x,\sigma)]+\mathbf{e}[{\epsilon^a}_b(\omega\, e^b)(x,\sigma)]+\boldsymbol{\omega}[\extd \omega(x,\sigma)]+ \boldsymbol{\omega}[\frac{1}{2}\epsilon_{ab}\big(\int_0^\sigma (e^a e^b)(x,\sigma^\p)d \sigma^\p\big)]\,.
\end{align}
We also introduce the Level-$\infty$ dilaton by extending \eqref{BFscalrn}
\begin{equation}\label{infdil}
    \mathcal{B}=\mathbf{Z}[Z^a(x,\sigma)]+ \mathbf{Y}[Y(x,\sigma)]\,,
\end{equation}
where we define
\begin{align}
  \mathbf{Z}[Z^a(x,\sigma)]=\sum^\infty_{k=0} \frac{\sigma^k }{k!}Z^a_{(k)}(x)P_a^{(k)}\,,\qquad\qquad
    \mathbf{Y}[Y(x,\sigma)]=\sum^\infty_{k=0}\frac{\sigma^k }{k!} Y_{(n)}(x)J^{(k)}\,.
\end{align}
\paragraph{Reduction to level-$N$.} The definitions \eqref{infgauge} and \eqref{infdil} reduce to \eqref{BFgaugen} and \eqref{BFscalrn} for the level-$N$ theory, if  both $\mathbf{e}$ and $\mathbf{Z}$ are kept  up to order $\sigma^{N-1}$ while $\boldsymbol{\omega}$ and $\mathbf{Y}$ are kept up to order $\sigma^N$. The gauge field $A$  in \eqref{BFgaugen} will be identified with $\omega_{(N)}$, and the dilaton $X$ in \eqref{BFscalrn} with $Y_{(N)}$.

\subsection{The level-\texorpdfstring{$\infty$}{infty-BF} BF-action}
To specify the BF-action, one needs to fix the pairing between various fields defined above. Using \eqref{Nbilin}, we have
\begin{align}
    \big\langle  \mathbf{Y}[Y(x,\sigma)],\boldsymbol{\omega}[\omega(x,\sigma])\big\rangle_{\Gamma_N}&=\sum_{k,m=0}^N\frac{1}{k!m!}Y_k(x)\omega_m(x)\sigma^{k+m}\delta_{k+m-N}=\frac{\sigma^N}{N!}(Yw)_N(x)\,.
\end{align}
The pairing of two functions $ \mathbf{Y}[Y(x,\sigma)]$ and $\boldsymbol{\omega}[\omega(x,\sigma)]$ using the bilinear form $\Gamma_N$  is given by the $N$th coefficient in the 
 Taylor series of the function $(Y\omega)(x,\sigma)$, which itself has $N+1$ terms when written in terms of the Taylor coefficients of $Y(x,\sigma)$ and $\omega(x,\sigma)$. It is thus suggestive to consider the bilinear form $\Gamma_\infty$ defined by the $N+1$-vector $\boldsymbol{\gamma}=(1,1,\cdots,1)$, and $N\to\infty$. The pairing acquires the simple form
\begin{equation}
      \big\langle  \mathbf{Y}[Y(x,\sigma)],\boldsymbol{\omega}[\omega(x,\sigma])\big\rangle_{{\Gamma}_\infty}=(Y\omega)(x,\sigma)\,.
\end{equation}
Similarly, for $\mathbf{Z}$ and $\mathbf{e}$ one obtains
\begin{equation}
        \big\langle \mathbf{Z}[Z^a(x,\sigma)],\mathbf{e}[e^a(x,\sigma)]\big\rangle_{{\Gamma}_\infty}=-(Z^ae_a)(x,\sigma)\,.  
\end{equation}
Now we are ready to propose the Level-$\infty$ BF-theory
\begin{equation}\label{infinite action}
    I^{(\infty)}[\mathcal{B},\mathcal{A};\sigma]=\int\Big\langle  \mathcal{B}[Z^a,Y]\cF[e^a,\omega]\Big\rangle_{{\Gamma}_\infty}\,,
\end{equation}
where the parameter $\sigma$ is fixed to a certain value in the convergence radius of the Taylor series, and the integration is performed on a two-dimensional manifold. Explicitly, the action is
\begin{equation}
      I^{(\infty)}[\mathcal{B},\cA;\sigma]=\int  \Big\{Y\big(\extd \omega+\frac{1}{2}\epsilon_{ab}\int_0^\sigma(e^ae^b)(\sigma^\p)d \sigma^\prime\big)+x^a\big(\extd e^a+{\epsilon^a}_b\omega e^b\big)\Big\}\,.
\end{equation}
Variation with respect to $Y(x,\sigma)$ and $Z^a(x,\sigma)$ gives the following field equations
\begin{align}
    &\extd \omega(x,\sigma)+\frac{1}{2}\epsilon_{ab}\int_0^\sigma(e^ae^b)(x,\sigma^\p)d \sigma^\prime=0\,,\label{infcur}\\
    &\extd e^a(x,\sigma)+{\epsilon^a}_b \omega(x,\sigma)e^b(x,\sigma)=0\,.\label{inftor}
\end{align}

\subsection*{Effective curvature}
$ I^{(\infty)}[\B,\cA;\sigma]$ is defined at a fixed $\sigma$. Equation \eqref{inftor} solves $w(x,\sigma)$ in terms of $e^a(x,\sigma)$, while by taking a derivative of the two sides in \eqref{infcur} with respect to $\sigma$ we have
\begin{equation}\label{infcur1}
    \extd \omega^\p(x,\sigma)+\frac{1}{2}\epsilon_{ab}(e^ae^b)(x,\sigma)=0\,.
\end{equation}
Suppose that we define an effective two dimensional metric $G_{\mu\nu}(x,\sigma)$ depending on the parameter $\sigma$ as
\begin{equation}
    G_{\mu\nu}(x,\sigma)=e^a_\mu(x,\sigma)e^b_\nu(x,\sigma)\eta_{ab}\,.
\end{equation}
Using the relation between the Ricci scalar and the spin connection in two dimensions, $R=2\star\extd\omega$ we can find the curvature of the metric $G_{\mu\nu}$
\begin{equation}
   R[G]=
  \frac{\int_0^\sigma \text{det}\,e(x,\sigma^\p)\extd\sigma^\p}{\text{det}\,e(x,\sigma)}=
   \frac{\int_0^\sigma \sqrt{G(x,\sigma^\p)}\extd\sigma^\p}{\sqrt{G(x,\sigma)}}\,.
\end{equation}
We can simplify this expression by exploiting the gauge symmetry of each spin-2 field $e^a_{(k)}$ such that
\begin{equation}
e_{(k=0)}=\begin{pmatrix}
    1& a(x) \\
       0  & 1
\end{pmatrix},\qquad\qquad e_{(k>0)}=\begin{pmatrix}
    1& a_{(k)}(x) \\
       0  & 0
\end{pmatrix}\,.
\end{equation}
In this gauge, the determinant of the effective metric is equal to one. It follows that
\begin{equation}
    R[G]=\sigma.
\end{equation}
Of course the final result must be independent of the gauge. We conclude that a stack of infinitely many spin-2 fields present in the theory \eqref{infinite action} give rise a curved background depending on the parameter $\sigma$. The effective geometry is a de Sitter space if $\sigma>0$ and an anti-de Sitter space if $\sigma<0$. Regarding the flat space limit, one can take the limit $\sigma\to0$ from some non-vanishing value. This limit effectively switches off all the fields in $\A$ except $e^a$ and $\omega$, and the translation generators effectively commute. The background geometry is flat in this case.

\section{Center of the 2D extended Poincar\'e group}\label{centerapp}
In this part we would like to find the center for symmetry groups we studied in the main text. To this end, we use the lemma to the Baker–Campbell–Hausdorff formula 
\begin{align}\label{BCH-formula}
    e^{X} e^{Y}= e^{(Y+\left[X,Y\right]+\frac{1}{2!}[X,[X,Y]]+\frac{1}{3!}[X,[X,[X,Y]]]+\cdots)} ~e^X,
\end{align}
where $X$ and $Y$ are elements in the Lie algebra of a Lie group. The group element $g=  e^{X}$ will belong to the center of the group iff for any arbitrary element $Y$ we have
\begin{align}\label{centercon}
    \left[X,Y\right]+\frac{1}{2!}[X,[X,Y]]+\frac{1}{3!}[X,[X,[X,Y]]]+\cdots=0\,.
\end{align}
This leads to constraints on parameters in the Lie algebra elements of $X$. 

\subsection{Center of the 2D Maxwell group}
First, we are going to explore the center of 2D Maxwell group at level-1. We denote the $X$ and $Y$ elements as
\begin{align}
    X&=\alpha J+\beta^+ P_{+}+\beta^- P_{-}+\alpha_{1} J^{(1)}\,,\\
    Y&=aJ+b^+ P_{+}+b^- P_{-}+a_{1}J^{(1)}\,.\nn
\end{align}
Utilizing the 2D Maxwell algebra commutation relations \eqref{Maxwel1} in \eqref{BCH-formula} we obtain
\begin{align}
     e^{X} e^{Y}=& e^{\left(Y+A^+ P_{+} + A^- P_{-} + C_{1} J^{(1)} \right)} ~e^X\,,
\end{align}
where the coefficients are given by\footnote{The case $\alpha=0$ gives the trivial center $\beta_\pm=0$.}
\begin{align}
    A^\pm&=(\frac{\beta^\pm a}{\alpha}-b^\pm)(1-e^{\mp\alpha})\,,\nn\\
    C_{1}&=(\frac{\beta^+ b^-}{\alpha})(e^{\alpha}-1)+(\frac{\beta^- b^+}{\alpha})(e^{-\alpha}-1)+\frac{2\beta^+ \beta^- a}{\alpha^{2}}(1-\cosh\alpha).
\end{align}
 The  center condition \eqref{centercon} is satisfied for the Lie algebra element $X$ by the nontrivial solution to $A^\pm=C_{1}=0$ which in this case is 
 \begin{align}
 \alpha=\pm2\pi \,i\, n    \,,\qquad n\in \mathbb{N}\,.
 \end{align}
Other parameters in $X$ remain unconstrained.

 \subsection{Center of the extended Maxwell group}
  In the next step, we are going to explore the center of the group at level-2. We denote $X$ and $Y$ elements as\footnote{We considered $X$ as an element of the algebra in accordance to our gauge fixed boundary conditions \eqref{L2bc}.}
  \begin{align}\label{centercond2}
      X&=\alpha J+\beta^{+} P_{+}+\beta^{-} P_{-}+\beta^{+}_{1} P^{(1)}_{+}+\beta^{-}_{1} P^{(1)}_{-}\,,\\
      Y&=aJ+b^{+} P_{+}+b^{-} P_{-}+a_{1}J^{(1)}+b^{+}_{1}P^{(1)}_{+}+b^{-}_{1}P_{-}^{(1)}+a_{2}J^{(2)}\,.\nn
  \end{align}
  We explore constraints on the group element $X$ such that it is the   center of the extended Maxwell group. In general we have
  \begin{align}
     e^{X} e^{Y}= e^{\left(Y+A^{+}P_{+}+A^{-}P_{-}+C_{1}J^{(1)}+A_{1}^{+}P^{(1)}_{+}+A_{1}^{-}P^{(1)}_{-}+C_{2}J^{(2)}\right)} ~e^X,
\end{align}
 The commutators of extended Maxwell algebra \eqref{level2alglc} and the BCH lemma in \eqref{BCH-formula} leads to  the coefficients given as 
    \begin{align}\label{constraint-center-extended}
    A^\pm&=(\frac{\beta^\pm a}{\alpha}-b^\pm)(1-e^{\mp\alpha})\,,\nn\\
      C_{1}&=\frac{\beta^{+} b^{-}}{\alpha}(e^{\alpha}-1)+\frac{\beta^{-} b^{+}}{\alpha}(e^{-\alpha}-1)+\frac{2\beta^{+} \beta^{-} a}{\alpha^2}(\cosh \alpha-1)\,,\nn\\
        A^\pm_{1}&=(\frac{\beta^\pm_{1} a_{1}}{\alpha}+\frac{a \beta^{\pm}_{1}}{\alpha}-b^\pm_{1})(1-e^{\mp\alpha})+\frac{b^{\mp}{\beta^{\pm}}^2}{\alpha^2} (\cosh\alpha-1)+\frac{b^{\pm} \beta^{-} \beta^{+}}{\alpha^2}(e^{\mp\alpha}-1+\alpha e^{\mp\alpha}),\nn\\
      & +\frac{a {\beta^{\pm}}^2\beta^{\mp} }{2\alpha^3}(\mp2\alpha e^{\mp\alpha}-3e^{\mp\alpha}-e^{\pm\alpha}+4),\nn\\
      C_{2}&=(\frac{b^{+}_{1}\beta^{-}}{\alpha}+\frac{b^{+} \beta^{-}_{1}}{\alpha})(e^{-\alpha}-1 )+(\frac{b^{-}_{1} \beta^{+}}{\alpha}+\frac{b^{-} \beta^{+}_{1}}{\alpha})(e^{\alpha}-1 )+\frac{4a{\beta^{+}}^2 {\beta^{-}}^2}{\alpha^4}(\frac{\alpha}{2}\sinh\alpha-\cosh\alpha+1)\nn\\
      &+(\frac{2a_1 \beta^{+} \beta^{-}}{\alpha^2}+\frac{2a \beta^{-} \beta^{+}_{1}}{\alpha^2}+\frac{2a \beta^{+} \beta^{-}_{1}}{\alpha^2})(1-\cosh\alpha)-\frac{b^{-}{\beta^{+}}^2 \beta^{-}}{2\alpha^3}(2\alpha e^{\alpha}-3e^{\alpha}-e^{-\alpha}+4)\nn\\
      &-\frac{b^{+} \beta^{+} {\beta^{-}}^2}{2\alpha^3}(-2\alpha e^{-\alpha}-3e^{-\alpha}-e^{\alpha}+4)\,,
  \end{align}
  The center condition \eqref{centercon} is satisfied by finding a non-trivial solution to $A^\pm=C_{1}=A^\pm_{1}=C_{2}=0$ in \eqref{centercond2}. Again it necessitate to have
  \begin{align}\label{alphcenter}
      \alpha=\pm2\pi\,i\, n\,,\qquad n\in\mathbb{N}\,.
  \end{align}
However there remains three terms in $A^\pm_{1}$ and $C_{2}$ in \eqref{constraint-center-extended} which do not become zero after imposing \eqref{alphcenter}. These  terms which should become independently  zero are 
  \begin{align}
  0&=  \frac{b^{+}\beta^{-} \beta^{+}}{\alpha}-\frac{a {\beta^{+}}^2 \beta^{-}}{\alpha^2}\,,\nn\\
 0&=  \frac{b^{-} \beta^{-} \beta^{+}}{\alpha}+\frac{a \beta^{+} {\beta^{-}}^2}{\alpha^2}\,,\nn\\
 0&=\frac{b^{+} \beta^{+} {\beta^{-}}^2}{\alpha^2} e^{-2\alpha}-\frac{b^{-}{\beta^{+}}^2 \beta^{-}}{\alpha^2}\, .
  \end{align}
 First of all these extra conditions show that $\beta^\pm_1$ are unconstrained. 
Since we do not want to put any constraint on the arbitrary element $Y$, from the last equality we have 
\begin{align}
    \beta^{+} \beta^{-}=0\,,
\end{align}
which solves the other two conditions as well. Without loss of generality, we can make the choice $\beta^{-}=1$ which is consistent with our boundary conditions at level-2 \eqref{L2bc}.

\addcontentsline{toc}{section}{References}
\bibliographystyle{fullsort.bst}
\bibliography{reference}

\providecommand{\href}[2]{#2}\begingroup\raggedright\begin{thebibliography}{10}

\bibitem{Almheiri:2014cka}
A.~Almheiri and J.~Polchinski, ``{Models of AdS$_{2}$ backreaction and
  holography},'' {\em JHEP} {\bf 11} (2015) 014,
\href{http://www.arXiv.org/abs/1402.6334}{{\tt 1402.6334}}.

\bibitem{Maldacena:2016upp}
J.~Maldacena, D.~Stanford, and Z.~Yang, ``{Conformal symmetry and its breaking
  in two dimensional Nearly Anti-de-Sitter space},'' {\em PTEP} {\bf 2016}
  (2016), no.~12, 12C104,
\href{http://www.arXiv.org/abs/1606.01857}{{\tt 1606.01857}}.

\bibitem{Jensen:2016pah}
K.~Jensen, ``{Chaos in AdS$_2$ Holography},'' {\em Phys. Rev. Lett.} {\bf 117}
  (2016), no.~11, 111601,
\href{http://www.arXiv.org/abs/1605.06098}{{\tt 1605.06098}}.

\bibitem{Engelsoy:2016xyb}
J.~Engels{\"o}y, T.~G. Mertens, and H.~Verlinde, ``{An investigation of
  AdS$_{2}$ backreaction and holography},'' {\em JHEP} {\bf 07} (2016) 139,
\href{http://www.arXiv.org/abs/1606.03438}{{\tt 1606.03438}}.

\bibitem{Saad:2018bqo}
P.~Saad, S.~H. Shenker, and D.~Stanford, ``{A semiclassical ramp in SYK and in
  gravity},'' \href{http://www.arXiv.org/abs/1806.06840}{{\tt 1806.06840}}.

\bibitem{Saad:2019lba}
P.~Saad, S.~H. Shenker, and D.~Stanford, ``{JT gravity as a matrix integral},''
  \href{http://www.arXiv.org/abs/1903.11115}{{\tt 1903.11115}}.

\bibitem{Kitaev:15ur}
A.~Kitaev, ``{A simple model of quantum holography}.'' KITP strings seminars,
  April/May 2015,
  \href{http://online.kitp.ucsb.edu/online/entangled15/}{http://online.kitp.ucsb.edu/online/entangled15/}
  and
  \href{http://online.kitp.ucsb.edu/online/entangled15/kitaev2/}{http://online.kitp.ucsb.edu/online/entangled15/kitaev2/}.

\bibitem{Sachdev:1992fk}
S.~Sachdev and J.~Ye, ``{Gapless spin fluid ground state in a random, quantum
  Heisenberg magnet},'' {\em Phys. Rev. Lett.} {\bf 70} (1993) 3339,
\href{http://www.arXiv.org/abs/cond-mat/9212030}{{\tt cond-mat/9212030}}.

\bibitem{Sachdev:2010um}
S.~Sachdev, ``{Holographic metals and the fractionalized Fermi liquid},'' {\em
  Phys. Rev. Lett.} {\bf 105} (2010) 151602,
\href{http://www.arXiv.org/abs/1006.3794}{{\tt 1006.3794}}.

\bibitem{Kitaev:2017awl}
A.~Kitaev and S.~J. Suh, ``{The soft mode in the Sachdev-Ye-Kitaev model and
  its gravity dual},'' {\em JHEP} {\bf 05} (2018) 183,
  \href{http://www.arXiv.org/abs/1711.08467}{{\tt 1711.08467}}.

\bibitem{DHoker:1982wmk}
E.~D'Hoker and R.~Jackiw, ``{Liouville Field Theory},'' {\em Phys. Rev. D} {\bf
  26} (1982) 3517.

\bibitem{Teitelboim:1983ux}
C.~Teitelboim, ``Gravitation and {H}amiltonian structure in two space-time
  dimensions,'' {\em Phys. Lett.} {\bf B126} (1983)
41.

\bibitem{Jackiw:1984je}
R.~Jackiw, ``{Lower Dimensional Gravity},'' {\em Nucl. Phys. B} {\bf 252}
  (1985) 343--356.

\bibitem{Jackiw:1984}
R.~Jackiw, ``{Liouville field theory: A two-dimensional model for gravity?},''
  in {\em Quantum Theory Of Gravity}, S.~Christensen, ed., pp.~403--420.
\newblock Adam Hilger, Bristol, 1984.

\bibitem{Teitelboim:1984}
C.~Teitelboim, ``{The {H}amiltonian structure of two-dimensional space-time and
  its relation with the conformal anomaly},'' in {\em Quantum Theory Of
  Gravity}, S.~Christensen, ed., pp.~327--344.
\newblock Adam Hilger, Bristol, 1984.

\bibitem{Maldacena:2016hyu}
J.~Maldacena and D.~Stanford, ``{Remarks on the Sachdev-Ye-Kitaev model},''
  {\em Phys. Rev.} {\bf D94} (2016), no.~10, 106002,
\href{http://www.arXiv.org/abs/1604.07818}{{\tt 1604.07818}}.

\bibitem{Mandal:1991tz}
G.~Mandal, A.~M. Sengupta, and S.~R. Wadia, ``Classical solutions of
  two-dimensional string theory,'' {\em Mod. Phys. Lett.} {\bf A6} (1991)
1685--1692.

\bibitem{Witten:1991yr}
E.~Witten, ``On string theory and black holes,'' {\em Phys. Rev.} {\bf D44}
  (1991)
314--324.

\bibitem{Nayak:2018qej}
P.~Nayak, A.~Shukla, R.~M. Soni, S.~P. Trivedi, and V.~Vishal, ``{On the
  Dynamics of Near-Extremal Black Holes},'' {\em JHEP} {\bf 09} (2018) 048,
  \href{http://www.arXiv.org/abs/1802.09547}{{\tt 1802.09547}}.

\bibitem{Grumiller:2015vaa}
D.~Grumiller, J.~Salzer, and D.~Vassilevich, ``{AdS$_{2}$ holography is
  (non-)trivial for (non-)constant dilaton},'' {\em JHEP} {\bf 12} (2015) 015,
\href{http://www.arXiv.org/abs/1509.08486}{{\tt 1509.08486}}.

\bibitem{Cvetic:2016eiv}
M.~Cveti\v{c} and I.~Papadimitriou, ``{AdS$_{2}$ holographic dictionary},''
  {\em JHEP} {\bf 12} (2016) 008,
  \href{http://www.arXiv.org/abs/1608.07018}{{\tt 1608.07018}}.
[Erratum: JHEP01,120(2017)].

\bibitem{Davison:2016ngz}
R.~A. Davison, W.~Fu, A.~Georges, Y.~Gu, K.~Jensen, and S.~Sachdev,
  ``{Thermoelectric transport in disordered metals without quasiparticles: The
  Sachdev-Ye-Kitaev models and holography},'' {\em Phys. Rev.} {\bf B95}
  (2017), no.~15, 155131,
\href{http://www.arXiv.org/abs/1612.00849}{{\tt 1612.00849}}.

\bibitem{Grumiller:2017qao}
D.~Grumiller, R.~McNees, J.~Salzer, C.~Valc\'{a}rcel, and D.~Vassilevich,
  ``{Menagerie of AdS$_{2}$ boundary conditions},'' {\em JHEP} {\bf 10} (2017)
  203,
\href{http://www.arXiv.org/abs/1708.08471}{{\tt 1708.08471}}.

\bibitem{Mertens:2018fds}
T.~G. Mertens, ``{The Schwarzian theory \textemdash{} origins},'' {\em JHEP}
  {\bf 05} (2018) 036, \href{http://www.arXiv.org/abs/1801.09605}{{\tt
  1801.09605}}.

\bibitem{Alekseev:1988ce}
A.~Alekseev and S.~L. Shatashvili, ``{Path Integral Quantization of the
  Coadjoint Orbits of the Virasoro Group and 2D Gravity},'' {\em Nucl. Phys.}
  {\bf B323} (1989)
719--733.

\bibitem{Witten:1987ty}
E.~Witten, ``{Coadjoint Orbits of the Virasoro Group},'' {\em Commun. Math.
  Phys.} {\bf 114} (1988)
1.

\bibitem{Afshar:2019tvp}
H.~R. Afshar, ``{Warped Schwarzian theory},'' {\em JHEP} {\bf 02} (2020) 126,
\href{http://www.arXiv.org/abs/1908.08089}{{\tt 1908.08089}}.

\bibitem{Afshar:2019axx}
H.~Afshar, H.~A. Gonz\'alez, D.~Grumiller, and D.~Vassilevich, ``{Flat space
  holography and the complex Sachdev-Ye-Kitaev model},'' {\em Phys. Rev. D}
  {\bf 101} (2020), no.~8, 086024,
  \href{http://www.arXiv.org/abs/1911.05739}{{\tt 1911.05739}}.

\bibitem{Cangemi:1992bj}
D.~Cangemi and R.~Jackiw, ``Gauge invariant formulations of lineal gravity,''
  {\em Phys. Rev. Lett.} {\bf 69} (1992) 233--236,
\href{http://arXiv.org/abs/hep-th/9203056}{{\tt hep-th/9203056}}.

\bibitem{Callan:1992rs}
C.~G. Callan, Jr., S.~B. Giddings, J.~A. Harvey, and A.~Strominger,
  ``Evanescent black holes,'' {\em Phys. Rev.} {\bf D45} (1992) 1005--1009,
\href{http://www.arXiv.org/abs/hep-th/9111056}{{\tt hep-th/9111056}}.

\bibitem{Strominger:1994tn}
A.~Strominger, ``Les {H}ouches lectures on black holes,''
  \href{http://www.arXiv.org/abs/arXiv:hep-th/9501071}{{\tt
  arXiv:hep-th/9501071}}.
Talk given at {NATO} Advanced Study Institute.

\bibitem{Grumiller:2002nm}
D.~Grumiller, W.~Kummer, and D.~V. Vassilevich, ``Dilaton gravity in two
  dimensions,'' {\em Phys. Rept.} {\bf 369} (2002) 327--429,
\href{http://arXiv.org/abs/hep-th/0204253}{{\tt hep-th/0204253}}.

\bibitem{Nojiri:2000ja}
S.~Nojiri and S.~D. Odintsov, ``Quantum dilatonic gravity in d = 2, 4 and 5
  dimensions,'' {\em Int. J. Mod. Phys.} {\bf A16} (2001) 1015--1108,
\href{http://arXiv.org/abs/hep-th/0009202}{{\tt hep-th/0009202}}.

\bibitem{Dubovsky:2017cnj}
S.~Dubovsky, V.~Gorbenko, and M.~Mirbabayi, ``{Asymptotic fragility, near
  AdS$_{2}$ holography and $ T\overline{T} $},'' {\em JHEP} {\bf 09} (2017)
  136, \href{http://www.arXiv.org/abs/1706.06604}{{\tt 1706.06604}}.

\bibitem{Fitkevich:2020okl}
M.~Fitkevich, D.~Levkov, and Y.~Zenkevich, ``{Dilaton gravity with a boundary:
  from unitarity to black hole evaporation},'' {\em JHEP} {\bf 20} (2020) 184,
  \href{http://www.arXiv.org/abs/2004.13745}{{\tt 2004.13745}}.

\bibitem{Detournay:2012pc}
S.~Detournay, T.~Hartman, and D.~M. Hofman, ``{Warped Conformal Field
  Theory},'' {\em Phys.Rev.} {\bf D86} (2012) 124018,
\href{http://www.arXiv.org/abs/1210.0539}{{\tt 1210.0539}}.

\bibitem{Afshar:2015wjm}
H.~Afshar, S.~Detournay, D.~Grumiller, and B.~Oblak, ``{Near-Horizon Geometry
  and Warped Conformal Symmetry},'' {\em JHEP} {\bf 03} (2016) 187,
\href{http://www.arXiv.org/abs/1512.08233}{{\tt 1512.08233}}.

\bibitem{Barnich:2006av}
G.~Barnich and G.~Comp{\`e}re, ``{Classical central extension for asymptotic
  symmetries at null infinity in three spacetime dimensions},'' {\em
  Class.Quant.Grav.} {\bf 24} (2007) F15--F23,
\href{http://www.arXiv.org/abs/gr-qc/0610130}{{\tt gr-qc/0610130}}.

\bibitem{Barnich:2009se}
G.~Barnich and C.~Troessaert, ``{Symmetries of asymptotically flat 4
  dimensional spacetimes at null infinity revisited},'' {\em Phys. Rev. Lett.}
  {\bf 105} (2010) 111103,
\href{http://www.arXiv.org/abs/0909.2617}{{\tt 0909.2617}}.

\bibitem{Bagchi:2010eg}
A.~Bagchi, ``{Correspondence between Asymptotically Flat Spacetimes and
  Nonrelativistic Conformal Field Theories},'' {\em Phys. Rev. Lett.} {\bf 105}
  (2010) 171601,
\href{http://www.arXiv.org/abs/1006.3354}{{\tt 1006.3354}}.

\bibitem{Barnich:2011mi}
G.~Barnich and C.~Troessaert, ``{BMS charge algebra},'' {\em JHEP} {\bf 1112}
  (2011) 105,
\href{http://www.arXiv.org/abs/1106.0213}{{\tt 1106.0213}}.

\bibitem{Barnich:2013axa}
G.~Barnich and C.~Troessaert, ``{Comments on holographic current algebras and
  asymptotically flat four dimensional spacetimes at null infinity},'' {\em
  JHEP} {\bf 1311} (2013) 003,
\href{http://www.arXiv.org/abs/1309.0794}{{\tt 1309.0794}}.

\bibitem{Oblak:2016eij}
B.~Oblak, ``{BMS Particles in Three Dimensions},''
\href{http://www.arXiv.org/abs/1610.08526}{{\tt 1610.08526}}.

\bibitem{Afshar:2013vka}
H.~Afshar, A.~Bagchi, R.~Fareghbal, D.~Grumiller, and J.~Rosseel, ``{Higher
  spin theory in 3-dimensional flat space},'' {\em Phys.Rev.Lett.} {\bf 111}
  (2013) 121603,
\href{http://www.arXiv.org/abs/1307.4768}{{\tt 1307.4768}}.

\bibitem{Safari:2019zmc}
H.~Safari and M.~Sheikh-Jabbari, ``{BMS$_{4}$ algebra, its stability and
  deformations},'' {\em JHEP} {\bf 04} (2019) 068,
  \href{http://www.arXiv.org/abs/1902.03260}{{\tt 1902.03260}}.

\bibitem{Parsa:2018kys}
A.~Farahmand~Parsa, H.~Safari, and M.~Sheikh-Jabbari, ``{On Rigidity of 3d
  Asymptotic Symmetry Algebras},'' {\em JHEP} {\bf 03} (2019) 143,
  \href{http://www.arXiv.org/abs/1809.08209}{{\tt 1809.08209}}.

\bibitem{Safari:2020pje}
H.~Safari, ``{Deformation of Asymptotic Symmetry Algebras and Their Physical
  Realizations},'' other thesis, 11, 2020.

\bibitem{Fukuyama:1985gg}
T.~Fukuyama and K.~Kamimura, ``{Gauge Theory of Two-dimensional Gravity},''
  {\em Phys. Lett. B} {\bf 160} (1985) 259--262.

\bibitem{Isler:1989hq}
K.~Isler and C.~A. Trugenberger, ``A gauge theory of two-dimensional quantum
  gravity,'' {\em Phys. Rev. Lett.} {\bf 63} (1989)
834.

\bibitem{Schaller:1994es}
P.~Schaller and T.~Strobl, ``Poisson structure induced (topological) field
  theories,'' {\em Mod. Phys. Lett.} {\bf A9} (1994) 3129--3136,
\href{http://arXiv.org/abs/hep-th/9405110}{{\tt hep-th/9405110}}.

\bibitem{Fradkin:1989uh}
E.~S. Fradkin and V.~Y. Linetsky, ``{Higher Spin Symmetry in One-dimension and
  Two-dimensions. 1.},'' {\em Mod. Phys. Lett. A} {\bf 4} (1989) 2635--2647.

\bibitem{Bengtsson:1986zm}
A.~K.~H. Bengtsson and I.~Bengtsson, ``{Higher 'Spins' in One and Two
  Space-time Dimensions},'' {\em Phys. Lett. B} {\bf 174} (1986) 294--300.

\bibitem{Vasiliev:1995dn}
M.~A. Vasiliev, ``{Higher spin gauge theories in four-dimensions,
  three-dimensions, and two-dimensions},'' {\em Int. J. Mod. Phys. D} {\bf 5}
  (1996) 763--797, \href{http://www.arXiv.org/abs/hep-th/9611024}{{\tt
  hep-th/9611024}}.

\bibitem{Alkalaev:2013fsa}
K.~Alkalaev, ``{On higher spin extension of the Jackiw-Teitelboim gravity
  model},''
\href{http://www.arXiv.org/abs/1311.5119}{{\tt 1311.5119}}.

\bibitem{Grumiller:2013swa}
D.~Grumiller, M.~Leston, and D.~Vassilevich, ``{Anti-de Sitter holography for
  gravity and higher spin theories in two dimensions},'' {\em Phys. Rev. D}
  {\bf 89} (2014), no.~4, 044001,
  \href{http://www.arXiv.org/abs/1311.7413}{{\tt 1311.7413}}.

\bibitem{Gonzalez:2018enk}
H.~A. Gonz{\'a}lez, D.~Grumiller, and J.~Salzer, ``{Towards a bulk description
  of higher spin SYK},'' {\em JHEP} {\bf 05} (2018) 083,
\href{http://www.arXiv.org/abs/1802.01562}{{\tt 1802.01562}}.

\bibitem{Alkalaev:2019xuv}
K.~Alkalaev and X.~Bekaert, ``{Towards higher-spin AdS$_2$/CFT$_1$
  holography},'' {\em JHEP} {\bf 04} (2020) 206,
  \href{http://www.arXiv.org/abs/1911.13212}{{\tt 1911.13212}}.

\bibitem{Alkalaev:2020kut}
K.~Alkalaev and X.~Bekaert, ``{On BF-type higher-spin actions in two
  dimensions},'' {\em JHEP} {\bf 05} (2020) 158,
  \href{http://www.arXiv.org/abs/2002.02387}{{\tt 2002.02387}}.

\bibitem{Grumiller:2020elf}
D.~Grumiller, J.~Hartong, S.~Prohazka, and J.~Salzer, ``{Limits of JT
  gravity},'' \href{http://www.arXiv.org/abs/2011.13870}{{\tt 2011.13870}}.

\bibitem{Gomis:2020wxp}
J.~Gomis, D.~Hidalgo, and P.~Salgado-Rebolledo, ``{Non-relativistic and
  Carrollian limits of Jackiw-Teitelboim gravity},''
  \href{http://www.arXiv.org/abs/2011.15053}{{\tt 2011.15053}}.

\bibitem{Hansen:2020hrs}
D.~Hansen, Y.~Jiang, and J.~Xu, ``{Geometrizing non-relativistic bilinear
  deformations},'' \href{http://www.arXiv.org/abs/2012.12290}{{\tt
  2012.12290}}.

\bibitem{Khasanov:2011jr}
O.~Khasanov and S.~Kuperstein, ``{(In)finite extensions of algebras from their
  Inonu-Wigner contractions},'' {\em J. Phys. A} {\bf 44} (2011) 475202,
  \href{http://www.arXiv.org/abs/1103.3447}{{\tt 1103.3447}}.

\bibitem{Verlinde:1991rf}
H.~Verlinde, ``Black holes and strings in two dimensions,'' in {\em Trieste
  Spring School on Strings and Quantum Gravity}, pp.~178--207.
\newblock April, 1991.
\newblock the same lectures were given at MGVI in Japan, June, 1991.

\bibitem{Nappi:1993ie}
C.~R. Nappi and E.~Witten, ``{A WZW model based on a nonsemisimple group},''
  {\em Phys. Rev. Lett.} {\bf 71} (1993) 3751--3753,
  \href{http://www.arXiv.org/abs/hep-th/9310112}{{\tt hep-th/9310112}}.

\bibitem{Stanford:2017thb}
D.~Stanford and E.~Witten, ``{Fermionic Localization of the Schwarzian
  Theory},'' {\em JHEP} {\bf 10} (2017) 008,
  \href{http://www.arXiv.org/abs/1703.04612}{{\tt 1703.04612}}.

\bibitem{Witten:1988xj}
E.~Witten, ``Topological sigma models,'' {\em Commun. Math. Phys.} {\bf 118}
  (1988)
411.

\bibitem{Campoleoni:2010zq}
A.~Campoleoni, S.~Fredenhagen, S.~Pfenninger, and S.~Theisen, ``{Asymptotic
  symmetries of three-dimensional gravity coupled to higher-spin fields},''
  {\em JHEP} {\bf 1011} (2010) 007,
  \href{http://www.arXiv.org/abs/1008.4744}{{\tt 1008.4744}}.

\bibitem{Gonzalez:2013oaa}
H.~A. Gonzalez, J.~Matulich, M.~Pino, and R.~Troncoso, ``{Asymptotically flat
  spacetimes in three-dimensional higher spin gravity},'' {\em JHEP} {\bf 1309}
  (2013) 016,
\href{http://www.arXiv.org/abs/1307.5651}{{\tt 1307.5651}}.

\bibitem{Godet:2020xpk}
V.~Godet and C.~Marteau, ``{New boundary conditions for AdS$_{2}$},'' {\em
  JHEP} {\bf 12} (2020) 020, \href{http://www.arXiv.org/abs/2005.08999}{{\tt
  2005.08999}}.

\bibitem{Bunster:2014mua}
C.~Bunster, M.~Henneaux, A.~Perez, D.~Tempo, and R.~Troncoso, ``{Generalized
  Black Holes in Three-dimensional Spacetime},'' {\em JHEP} {\bf 1405} (2014)
  031,
\href{http://www.arXiv.org/abs/1404.3305}{{\tt 1404.3305}}.

\bibitem{deBoer:1993iz}
J.~de~Boer and T.~Tjin, ``{The Relation between quantum W algebras and Lie
  algebras},'' {\em Commun. Math. Phys.} {\bf 160} (1994) 317--332,
  \href{http://www.arXiv.org/abs/hep-th/9302006}{{\tt hep-th/9302006}}.

\bibitem{Campoleoni:2011hg}
A.~Campoleoni, S.~Fredenhagen, and S.~Pfenninger, ``{Asymptotic W-symmetries in
  three-dimensional higher-spin gauge theories},'' {\em JHEP} {\bf 1109} (2011)
  113,
\href{http://www.arXiv.org/abs/1107.0290}{{\tt 1107.0290}}.

\bibitem{Gutperle:2011kf}
M.~Gutperle and P.~Kraus, ``{Higher Spin Black Holes},'' {\em JHEP} {\bf 1105}
  (2011) 022,
\href{http://www.arXiv.org/abs/1103.4304}{{\tt 1103.4304}}.

\bibitem{Gary:2012ms}
M.~Gary, D.~Grumiller, and R.~Rashkov, ``{Towards non-AdS holography in
  3-dimensional higher spin gravity},'' {\em JHEP} {\bf 1203} (2012) 022,
\href{http://www.arXiv.org/abs/1201.0013}{{\tt 1201.0013}}.

\bibitem{Afshar:2012nk}
H.~Afshar, M.~Gary, D.~Grumiller, R.~Rashkov, and M.~Riegler, ``{Non-AdS
  holography in 3-dimensional higher spin gravity - General recipe and
  example},'' {\em JHEP} {\bf 1211} (2012) 099,
\href{http://www.arXiv.org/abs/1209.2860}{{\tt 1209.2860}}.

\bibitem{Afshar:2012hc}
H.~Afshar, M.~Gary, D.~Grumiller, R.~Rashkov, and M.~Riegler, ``{Semi-classical
  unitarity in 3-dimensional higher-spin gravity for non-principal
  embeddings},'' {\em Class. Quant. Grav.} {\bf 30} (2012) 104004,
\href{http://www.arXiv.org/abs/1211.4454}{{\tt 1211.4454}}.

\bibitem{Castro:2011fm}
A.~Castro, E.~Hijano, A.~Lepage-Jutier, and A.~Maloney, ``{Black Holes and
  Singularity Resolution in Higher Spin Gravity},'' {\em JHEP} {\bf 1201}
  (2012) 031,
\href{http://www.arXiv.org/abs/1110.4117}{{\tt 1110.4117}}.

\bibitem{Perez:2012cf}
A.~Perez, D.~Tempo, and R.~Troncoso, ``{Higher spin gravity in 3D: black holes,
  global charges and thermodynamics},''
\href{http://www.arXiv.org/abs/1207.2844}{{\tt 1207.2844}}.

\bibitem{deBoer:2013gz}
J.~de~Boer and J.~I. Jottar, ``{Thermodynamics of higher spin black holes in
  $AdS_3$},'' {\em JHEP} {\bf 1401} (2014) 023,
\href{http://www.arXiv.org/abs/1302.0816}{{\tt 1302.0816}}.

\end{thebibliography}\endgroup

\end{document}